\documentclass{article}
\usepackage[T1]{fontenc}
\usepackage{graphicx} % Required for inserting images
\usepackage{amssymb}
\usepackage{amsthm}
\usepackage{comment}
\usepackage{stmaryrd}
\usepackage{mathtools}
\usepackage{graphicx}
\usepackage{tikz}
\usetikzlibrary{arrows, automata,  
                positioning, 
                quotes, calc}
\usepackage{xspace}

\usepackage{hyperref}

\newtheorem{definition}{Definition}[section]

\newtheorem{theorem}{Theorem}
\newtheorem{proposition}{Proposition}

\newtheorem{remark}{Remark}
\newtheorem{assumption}{Assumption}

\newcommand{\defined}[2]{{#1}\!\downarrow_{#2}}
\newcommand{\undefined}[2]{{#1}\!\ndownarrow_{#2}}
\newcommand{\SETL}{\textsc{setl}\xspace}

\newcommand{\denote}[2]{\llbracket{#1}\rrbracket{#2}}

\newcommand{\reduct}[2]{{#1}\!\!\upharpoonright_{#2}}

\newcommand*\circled[1]{\tikz[baseline=(char.base)]{
            \node[shape=circle,draw,inner sep=2pt] (char) {#1};}}

\title{FastSet: Parallel Claim Settlement}
\author{Xiaohong Chen$^1$, Grigore Rosu$^{1,2}$\\\normalsize Pi Squared Inc. (\texttt{http://pi2.network})$^1$\\\normalsize University of Illinois Urbana-Champaign$^2$}

\date{}

\begin{document}

\maketitle

%\begin{quote} \it
%     This paper is dedicated to Gul Agha, leader of the actor model of concurrency, who has taught generations of students that concurrent programming is hard only if you do not have the right model. 
%\end{quote}

\begin{abstract}
FastSet is a distributed protocol for decentralized finance and settlement, which is inspired from both actors and blockchains.
Account holders cooperate by making claims, which can include payments, holding and transferring assets, accessing and updating shared data, medical records, digital identity, and mathematical theorems, among others.
The claims are signed by their owners and are broadcast to a decentralized network of validators, which validate and settle them.
Validators replicate the global state of the accounts and need not communicate with each other.
In sharp contrast to blockchains, strong consistency is purposely given up as a requirement.
Yet, many if not most of the blockchain benefits are preserved, while capitalizing on actor's massive parallelism.
The protocol is proved to be correct, despite its massively parallel nature.
\end{abstract}

\setcounter{tocdepth}{2} % only show sections and subsections
\tableofcontents

\section{Introduction}

Blockchains are strongly consistent distributed systems in which account holders cooperate by making transactions.
Transactions are intents of actions, which need to be given a state to execute in order to materialize.
Transactions are signed by their owners and broadcast to a decentralized network of nodes, sometimes called miners or validators.
The nodes play dice in order to identify a proposer for the next block (or sequence) of transactions.
Nodes replicate the global state and follow a consensus protocol by which the proposed block is processed by all the nodes, thus materializing all the transactions in the block and each node updating its state accordingly.
In the end, the transactions submitted to the blockchain are totally ordered, and thus so are the states in between.

Strongly consistent distributed systems~\cite{lamport1978time,herlihy1990linearizability,lamport1998paxos,ongaro2014raft} ``behave like one computer'', which is the desirable property for blockchains, but are notorious for higher latency and lower performance and scalability---they require coordination and communication between nodes to maintain consistency.
At the other extreme stands the actor model~\cite{Hewitt1973,Agha1986}, aiming at highly parallel computing and massive concurrency.
Actors are the basic building blocks of concurrent computation and communicate only through messages.
%Each actor maintains a private state.
In response to a message it receives, an actor can modify its own private state, create more actors together with code governing their behavior, and send messages to other actors.

There is no doubt that blockchains play an important role in modern finance.
Also, blockchains have demonstrated, in our view irreversibly, that decentralization is not only possible in the world of digital assets, but also very much desirable.
Not only decentralization addresses the single point of failure, corruption, and censorship vulnerabilities, but equally importantly, through blockchains, it has lead to the Web3 philosophy and movement: users own their digital assets, from money to diplomas and medical records to pictures and messages, and they and only they can transfer them and decide who has access to what.
However, a major overhaul is needed in order to scale blockchains and achieve the level of high performance and low cost required by recent applications.
For example, few blockchains can consistently perform more than 1,000 transactions per second (TPS) and it takes seconds, sometimes minutes, to settle a transaction.
Metaphorically, because of the total order on transactions that they enforce by their nature, blockchains require the entire universe to squeeze through a narrow pipe.
Completely unrelated transactions are required to stay in line and wait to be sequenced in some order that the blockchain must globally choose when forming its next block.
This is clearly not scalable, even if all transactions are initiated by humans.
The situation is in fact much worse, because AI and AI agents doing transactions on humans' behalf are already here to stay and they require a payment system able to perform millions of TPS, most of which micro-transactions whose cost is expected to be negligible, in the order of fractions of a cent.
Blockchains cannot do this.
A massively parallel decentralized infrastructure for payments in particular and computing in general is required.

A series of papers before 2008 culminating with Bitcoin~\cite{nakamoto2008bitcoin}, have incrementally built a belief that a total order on transactions ought to be required in any distributed/decentralized payment system in order to avoid the infamous double spending attack: an account sending two concurrent transactions attempting to spend the same coin with two different recipients.
A total ordering enforced on transaction settlement indeed solves the double spending problem, but is it really necessary?
Recent works starting around 2019~\cite{Guerraoui2019,guerraoui2022consensus,Baudet2019FastPay} propose a radically different way to approach the problem, a truly concurrent approach where payments can be generated and settled in different orders by different nodes without breaking the semantics of payments.
The key insight of these works is that the order in which an account receives payments is irrelevant, and so is the order in which different accounts send payments---provided each account stays valid: no double spending and sufficient balance.
That is, as far as the nodes/validators are in agreement on the \textit{set} of locally valid transactions that took place in the system, consensus on a total order is unnecessary.
We believe that this apparently simple and innocent observation marks a crucial breakthrough moment in decentralized finance.
A moment where the tyranny of sequentiality is abolished and the door is open to innovations that will lead to the next generation of decentralized, yet truly concurrent, safe, and scalable infrastructure for digital assets.

Inspired by this recent work on weaker variants of consensus in the context of cryptocurrencies~\cite{Guerraoui2019,guerraoui2022consensus,Baudet2019FastPay}, as well as by the unbounded concurrent computing promise of the actor model~\cite{Hewitt1973,Agha1986}, in this paper we propose FastSet, a general-purpose distributed computing protocol that generalizes the payment system protocol FastPay~\cite{Baudet2019FastPay}.
FastSet performs nearly embarrassingly parallel settlement of arbitrary verifiable claims, including verifiable computations in arbitrary programming languages.
Specifically, FastSet allows a set of \textit{clients} to settle verifiable \textit{claims} consistently, using a set of \textit{validators}.
Clients broadcast their claims to all validators.
Validators validate the received claims and replicate the system state based on the order in which they receive the claims.
Importantly, similarly to FastPay but in sharp contrast to blockchains, the FastSet validators do not need to communicate with each other, nor directly achieve consensus on any values or orders or blocks or computations.

What makes the problem difficult is that claims can have side effects on the validators' states.
However, if claims issued by different clients are \textit{weakly independent}, a notion that generalizes the property of commutativity on payments~\cite{Guerraoui2019,guerraoui2022consensus,Baudet2019FastPay} discussed above to arbitrary state-effectful computations, then the validators' states will remain (strongly eventually) consistent in spite of the different orders in which they receive and process the claims.

Like in blockchains, FastSet accounts have a state (e.g., a balance) and are required to sign any claims they issue.
We present the protocol in Section~\ref{sec:fastset}, requiring only abstract notions of state, claim, and claim validity and effect.
How claims are generated, e.g., randomly by users or programmatically by contracts, is irrelevant for the core protocol and its correctness discussed in Section~\ref{sec:correctness}.
However, to make it practical, implementations of FastSet have to offer specific types of claims and specific ways to generate them.
In Section~\ref{sec:apps} we propose an actor-inspired language for generating claims, which we call the FastSet Language and abbreviate \SETL\footnote{Not to be confused with the SET Language \texttt{https://en.wikipedia.org/wiki/SETL}, invented in the late 1960s, based on the mathematical theory of sets.}, and pronounce ``settle''.
Some accounts are controlled by users, others by \SETL programs (or scripts, or contracts).
The difference is that the user-controlled accounts are free to issue any claims, including ones that create new accounts and interact with them, while the contract-based accounts can only issue claims as prescribed by their \SETL program/script/contract.

Like in the actor model, accounts are regarded as actors that communicate only with the actors they know about, and can create new accounts/actors and then communicate with them.
Since validators maintain replicas of the global system state, all the actors/accounts are also replicated on all validators.
This should not be regarded as a deviation from the underlying thesis of the actor model, where each actor is meant to be a separate process interacting concurrently with the other actors, but rather as a high-availability high-resilience decentralized implementation of an actor system.
Moreover, the actor model is particularly suitable for a language like \SETL for two additional reasons:
(1) each validator can itself be a high-performance concurrent system, which receives and processes potentially millions of claims per second, so \SETL must be a high-performance concurrent programming language; and (2) since each validator receives the claims in different orders, yet all of them (strongly eventually consistently) are expected to replicate the same state, \SETL concurrent programs must be easy to reason about, in particular to prove their determinism.

The rest of the paper is organized as follows. 
Section~\ref{sec:fastset} defines the FastSet Protocol.
Section~\ref{sec:formalization-correctness} proves the correctness of the protocol.
Section~\ref{sec:apps} proposes, through examples that are commonly encountered in Web3, an informal design of \SETL, a contract scripting language for FastSet.
Section~\ref{sec:conclusion} concludes the paper.
This paper also has two appendix sections. 
Appendix~\ref{sec:understanding-SETL} elaborates on the basic concepts that help the reader
to better understand the examples in Section~\ref{sec:apps}.
Appendix~\ref{sec:database} discusses some possible extensions of the examples.

We recommend the reader to first read Section~\ref{sec:fastset}, to understand how the FastSet protocol works.
Then readers can take one of the following two paths.
The theoretically inclined reader can dive into the mathematical formalization and the proofs of correctness of FastSet in Section~\ref{sec:formalization-correctness}.
The reader interested in applications that can take advantage of FastSet's massively parallel architecture can skim Section~\ref{sec:formalization-correctness} and skip directly to Section~\ref{sec:apps}.
If some examples are difficult to understand, the reader can check out the explanations in Appendix~\ref{sec:understanding-SETL}.

\section{The FastSet Protocol}
\label{sec:fastset}

In this section we describe the FastSet protocol, depicted in Figure~\ref{fig:fastset}.

\subsection{Participants}

FastSet involves two types of participants:
\begin{description}
    \item[Clients.] These are account/address holders, who can make \textit{claims}.
    They can be users, apps (contracts, web2, games), L1s, L2s, micro-chains, AI agents, service providers, execution engines/layers, provers (mathematical, ZK), TEEs, oracles, VRFs, AI compute/inferencers, indexers, history query providers, verifiers (execution, semantics and/or ZK proof based), fact claimers (KYC providers, digital identity providers, academic or medical records, etc.), token issuers (stablecoins, ERCs, NFTs, RWAs, etc.), etc.

    \item[Validators.] These process claims made by the clients, while at the same time maintaining consistency in the global knowledge about the overall system: balances of all tokens and additional data/state of all clients, global state of the entire protocol, like the set of all the claims that were settled, etc.
\end{description}
All participants possess a key pair consisting of a private signature key and the corresponding public verification key.
If $\textit{msg}$ is a message and $p$ is a participant, then $\langle \textit{msg} \rangle_p$ denotes the message signed by $p$, whose authenticity can be publicly checked.
Like in blockchains, in FastSet the public key of a client $a$ serves as the public account number, or the address of $a$.
Additional layers of privacy can be added if desired, but that is out of the scope of this paper.

\subsection{Claims and Weak Independence}
\label{sec:fs-wi}
\label{sec:crdt}

We define FastSet parametrically, on top of two important abstractions, \textit{claims} and their \textit{weak independence}, which we discuss next.

A \textit{claim} is any statement that is independently \textit{verifiable}.
To help the verifiers, the claim provider may associate additional evidence to the claim, such as a proof or a witness or even an authoritative signature; how claims provide evidence and how verifiers verify a claim is orthogonal to FastSet and is not our concern in this paper.
Here are some examples of claims:
``I am Joe Smith'' (a fact where the signer is important); ``Pythagoras theorem'' (a fact where the signer is less important); ``the price of gold is 100 USD'' (an oracle); ``the next random number is 17'' (a VRF); ``I want to buy a ticket to this concert'' (an intent); ``the result to your query is 42'' (an AI service provided); ``Python program \texttt{fibonacci} on input 10 evaluates to 55'' (verifiers may re-execute, or require a math or ZK proof based on Python formal semantics); ``my Angry Birds score is 739'' (requires 3rd party or ZK proof); ``my next move in this chess game with Alice is Nf3'' (modifying a shared storage location sequentially); ``I vote YES for that petition'' (modifying a shared storage location non-sequentially); ``I, Grigore, pay Alice 10 USD'' (a payment, modifying two storage locations).

For simplicity and generality, we only assume a global state in FastSet, which will be replicated in each validator.
In practice, each client will have their own reserved state space, but that stronger assumption is not needed to prove the correctness of FastSet; all we need is that claims issued by different clients are weakly independent, to be discussed shortly.
What is important is that a claim may or may not be valid in a given state (e.g., a payment is invalid when the sender's balance is not enough), and that the processing of a claim can and usually does change the state (e.g., even an otherwise side-effect-free claim may be counted, at a minimum).
We write $\defined{c}{s}$ whenever a claim $c$ is valid in a state $s$, and in that case $\denote{c}{s}$ denotes the state obtained after processing $c$ in $s$.
We extend these notations to sequences of claims $c_1\,c_2\,...\,c_k$ as expected:
$\defined{c_1\,c_2\,...\,c_k}{s}$ means that each claim in the sequence is valid in the state obtained after processing the previous ones, and $\denote{c_1\,c_2\,...\,c_k}{s}$ is the state obtained after processing the entire sequence.
The root of difficulty in FastSet, as well as in concurrent and distributed systems in general, is the fact that claim sequences issued by different interacting clients may arrive to validators interleaved in different ways, which may make their states diverge inconsistently.

Drawing inspiration from concurrency theory (e.g., Mazurkiewicz traces~\cite{Mazurkiewicz1987}), where two events are independent iff they can be processed in any order with the same result, we say that two claims $c$ and $c'$ are \textit{weakly independent}, written $c \parallel c'$, iff once each of them is independently valid, they can be processed in any order and the final result is the same:
$c \parallel c'$ iff for any state $s$, if $\defined{c}{s}$ and $\defined{c'}{s}$ then $\defined{c\,c'}{s}$, $\defined{c'\,c}{s}$, and $\denote{c\,c'}{s} = \denote{c'\,c}{s}$.
Therefore, we weaken the classic notion of independence by requiring it to hold only in those states in which both claims are already valid.
This assumption is critical for FastSet because it allows to handle payments and asset transfers as particular claims, and thus generalize FastPay~\cite{Baudet2019FastPay}. 
Here are some examples of weakly independent claims:
\begin{itemize}
    \item Totally (not weakly) independent claims whose validity has nothing to do with each other, such as: facts which are independently verifiable/true; state queries which are pure, that is, return results but have no side effects; state accesses, including writes/updates, but of disjoint storage locations.
    \item Commutative operations on a shared location: writing the same value, like in setting a one-way flag to "true"; adding numbers, positive only like in voting or both positive and negative like in reputation systems (e.g., the karma system of Reddit); multiplying numbers, like in aggregating signatures; adding elements to a set, like in signing petitions, bidding in an auction, or submitting (an intent to make) a transaction; adding elements to a canonical data-structure, like to a sorted list (which is re-sorted after each addition).
    \item Updates of CRDTs, abbreviating Commutative~\cite{Shapiro2007CRDTReport,Preguica2008CRDT} or Conflict-free~\cite{shapiro2011crdt} Replicated Data Types, which are data-types used in the context of distributed systems with the property that concurrent updates on them commute.
    CRDTs were used mainly in concurrent text editing, but the concept is general and includes many interesting examples, such as integer vectors, sets, maps, and even graphs (with some reasonable restrictions).
    \item Payments and, more generally, asset transfers from different clients.
\end{itemize}

Payments are, in particular, a very important use case of FastSet.
As critically observed by the authors of FastPay \cite{Baudet2019FastPay}, payments have the key property of commutativity on the recipient.
Specifically, if Charlie receives payments from both Alice and Bob, then Charlie will end up with the same balance/state no matter in what order the two payments from Alice and Bob are received.
And so will Alice and Bob.
It can be easily seen that any two payments made by distinct clients are weakly independent.
But they are \textit{not} necessarily independent with the standard notion of independence~\cite{Mazurkiewicz1987}.
For example, the claims ``Alice pays Bob 1 USD'' and ``Bob pays Charlie 1 USD'' are (weakly independent but) not independent: indeed, consider a state in which Alice has 1 USD and Bob has 0 USD.
The same state also demonstrates that two payments made by the same client are not necessarily weakly independent, e.g., ``Alice pays Bob 1 USD'' and ``Alice pays Charlie 1 USD''.
The weak independence of payments by different clients says that the overall state of the system will be the same independently of the order in which they are processed, provided that the initial state of the system allows for each payment to be independently made.

Non-examples of weak independence include claims acquiring the same mutex, or arbitrary accesses of a shared location where at least one is a write.
In particular, and this might look surprising at first sight, an exact balance claim is \textit{not} weakly independent with payment claims to that account.
For example, Alice's claim ``I have 20 USD'' is not weakly independent with payment claims made by others to Alice.
On the other hand, monotonic balance claims of the form ``I have at least 20 USD'' are weakly independent with payments made by others.
A challenge for FastSet implementations is to determine what constitutes claims on their network and how they want to enforce weak independence.

\subsection{Assumptions}
\label{sec:assumptions}

In its most basic form, FastSet's clients and validators form a bipartite graph, where each client is connected to each validator but the clients and, respectively, the validators, are not connected to each other.
Specifically, the only communication required by the basic protocol is for each client and each validator to be able to send signed messages to each other.
In practice, however, some clients will likely communicate with each other outside of the protocol in order to orchestrate and optimize their individual communication with the validators.
For example, a user's AI agent may require an approval from the user before buying a ticket; instead of communicating exclusively through claims and waiting for each of them to be settled by FastSet before proceeding to the next step, the user and the agent can both send their claims in parallel and settle the transaction faster and cheaper.
Similarly, in practice validators will likely have mechanisms to communicate with each other in order to synchronize their states faster than waiting for clients to send them the missing certificates.
However, such side communication mechanisms are outside the scope of this paper.

At a high-level, FastSet works as follows.
At any given moment, any client can broadcast to validators an ordered block of claims in a signed message.
Each validator checks the validity of each block of claims received from each client, and if everything checks the validator approves the block by signing the client's message and sending it back to the client without updating its state yet.
The client collects approvals from validators and once it reaches quorum it broadcasts a confirmation message to all validators.
Once a validator receives the confirmation it proceeds to updating its state.
Special care must be paid by the validator to order the state updates to avoid inconsistencies.
Eventually, all validators will consistently settle all the valid claims made by all the clients.

The validators do not question, nor check the intent behind or the client-specific semantics of the client's claims.
The validators only check the claims' validity and settle them if valid, operations which are expected to be very efficient; indeed, in practice these operations involve no computations other than updating client balances and storage locations with given values.
In other words, from validators' perspective the client claims are just simple commands that they must comply with after a few sanity checks.
Any computations using programs or smart contracts in various programming languages and VMs happen within the clients, using their resources and not validators'.
It is not validators' concern whether client's computations are adequate or correct according to client's intended semantics or specs or terms and conditions or whatever promises they may have made to their clients.
All the validators need to know is the block of claims that the client issues as a result of those potentially complex computations.
As an analogy, the validators play the role of an operating system (OS) in a computer, whose job is to ensure consistency across the various programs being executed; the programs are the clients of the OS and all they do from OS's perspective is to issue ordered blocks of commands (the claims).

FastSet therefore gives clients ultimate freedom in what claims they can issue.
Without any additional verification, it is quite possible that some clients may make mistakes, some of them even maliciously.
For example, the client may hold user assets, such as in the case of a bank, or a centralized exchange, or a blockchain.
That is, users may send their assets to the client in expectation of some services.
Without any additional verification, there is nothing to prevent the client from stealing users' assets.
Even without any evil intent, a client's account may be operated by a complex program which may have subtle errors (such as unintended non-determinism) and thus sometimes produce an incorrect sequence of claims.
Even without any program at all, a user client like Alice may mistype 11 instead of 1 when sending a payment to Bob, in which case additional verification, like a multi-sig, would have helped.

There is no silver-bullet solution to cover all types of claim verification that clients or applications need or will ever need.
To continue to give clients maximum flexibility, yet to allow the honest clients to verifiably prove themselves to their users, FastSet provides the capability for clients to set up verifiers and a quorum among those in order for its claims to be considered valid by the FastSet validators.
Verifiers are themselves clients on the network, whose role is to provide specialized verification services.
For example, a verifier can be specialized in verifying program executions in EVM (or Python, or Java, etc.) and clients, e.g., blockchains (or AI agents, or exchanges, etc.), can use it.
Such a verifier may re-execute the EVM program to check the result, or may verify a proof produced by an external compute, such as a ZK proof or a TEE proof or even just a signature by some trusted authority (e.g., a centralized exchange which offers services to its KYC-ed clients).
It is the client's full responsibility to decide what verifiers to include in their set and what quorum is sufficient.

Implementations of FastSet may optionally provide a simple hardwired programming or scripting language to generate sequences of claims.
Programs, or smart contracts, or scripts in this language would be executed by the validators on behalf of clients.
That is, instead of providing a block of claims to each validator, a client would provide a script that instructs validators how to generate the block of claims on their behalf, possibly using a client-provided input, at the cost of paying some gas fees to the validators. 
In this case, clients must pay special care to ensure that each validator deterministically generates the same claims in the same order.
Section~\ref{sec:apps} presents a candidate scripting language. 

To simplify the presentation of FastSet and to uniformly capture more use cases, we assume the existence of a proxy that facilitates the interaction between a client, its verifiers, and the validators.
The proxy is just a helper which only affects the liveness, but not the safety of the protocol.
All it does is to forward or broadcast messages signed by the various participants, and to blindly aggregate signatures ($\#\{\textit{sig}_1,\textit{sig}_2,...\}$).
Should the proxy fail to do its job for whatever reason, anybody can replace it and redo the tasks that were missed, including the client.
Proxy can be a client (the sender of the original message containing the claim sequence, or a different client such as a beneficiary or a verifier or a service provider), or a validator, or any combination of these for all or for any of the two services it provides, namely broadcasting client messages and aggregating validator signatures.
That is, there could be zero, one, or more proxies for each client or application deployed on FastSet.
If zero, then the client initiating the claims can do the job of the proxy, or any of the validators.

An honest validator always follows the FastSet protocol, while a faulty (or Byzantine) validator may deviate arbitrarily.
Assume $3f + 1$ validators, with a fixed unknown subset of at most $f$ Byzantine validators.
A \textit{quorum} is defined as any subset of $2f + 1$ validators.
A protocol message signed by a quorum of validators is \textit{certified} and called a \textit{certificate}.

\subsection{FastSet}

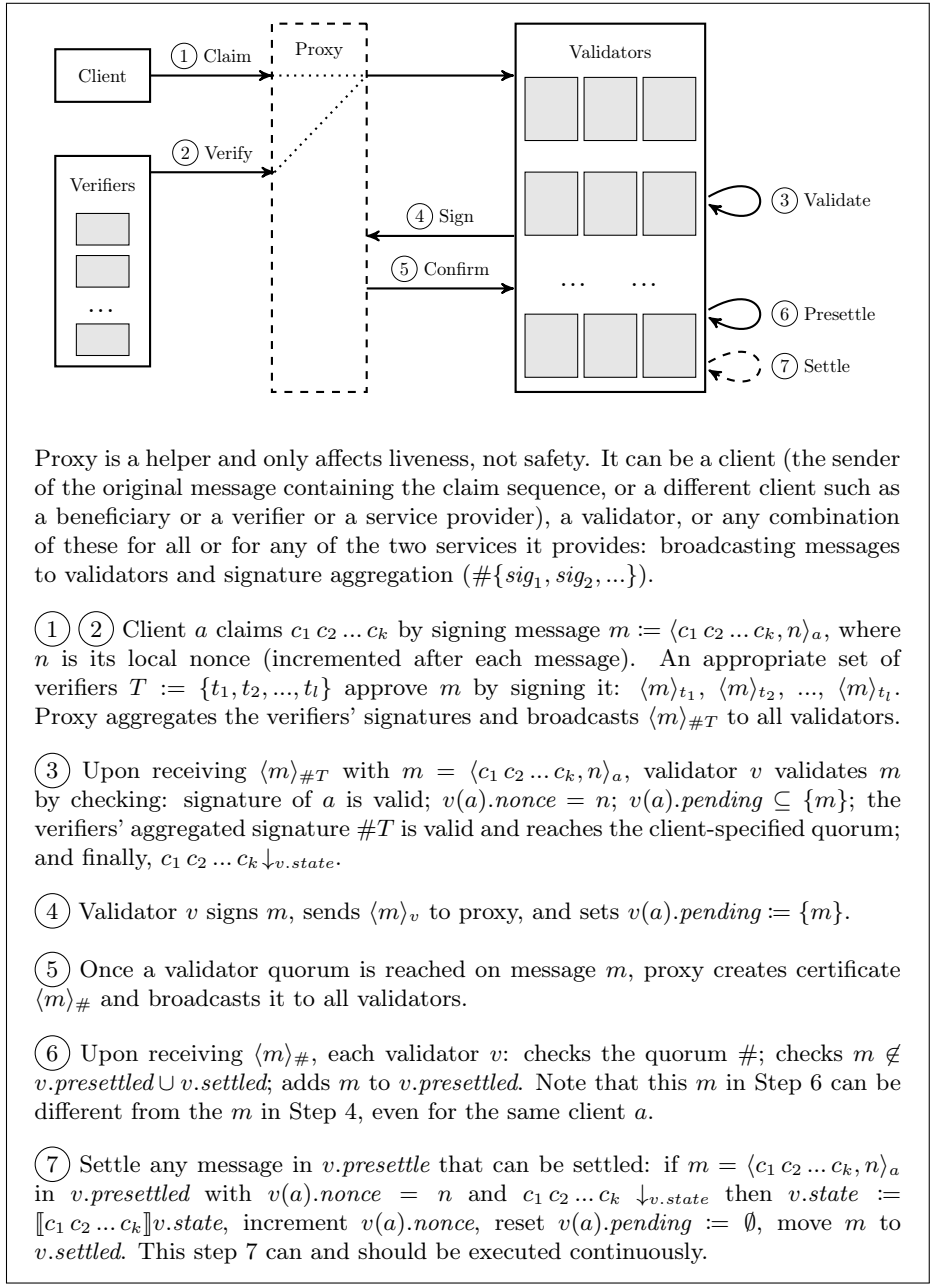
\begin{figure}
\setlength{\fboxsep}{2.5mm}
\fbox{
\begin{minipage}{.95\textwidth}
\begin{center}
\tikzset{
  verifier/.style={
    fill=gray!20,
    draw,
    rectangle,
    minimum width=1cm, 
    minimum height=.6cm,
  }
}

\tikzset{
  validator/.style={
    fill=gray!20,
    draw,
    rectangle,
    minimum width=1cm, 
    minimum height=1.2cm,
  }
}

\begin{tikzpicture}[scale=0.7, every node/.style={transform shape},every edge/.append style={thick}]
\node (client) [draw, thick, rectangle, minimum width=1.8cm, minimum height=1cm] {Client};
\node (verifiers) [draw, thick, rectangle, minimum width=1.8cm, minimum height=4cm, below=of client, label={[above=-8mm]above:Verifiers}] {};
\node[verifier,below=2.1cm of client] {};
\node[verifier,below=2.9cm of client] {};
\node[below=3.8 of client] {\bf\dots};
\node[verifier,below=4.2cm of client] {};

\node (proxy) [draw, dashed, thick, rectangle, minimum width=1.8cm, minimum height=7cm, label={[above=-8mm]above:Proxy}, right=3.2cm of client, anchor=north, yshift=1cm] {};

\node (validators) [draw, thick, rectangle, minimum width=3.6cm, minimum height=7cm, label={[above=-8mm]above:Validators}, right=2.8cm of proxy, anchor=west] {};
\node[validator,below=-6cm of validators] {};
\node[validator,below left=-6cm and -1.2cm of validators] {};
\node[validator,below right=-6cm and -1.2cm of validators] {};
\node[validator,below=-4.2cm of validators] {};
\node[validator,below left=-4.2cm and -1.2cm of validators] {};
\node[validator,below right=-4.2cm and -1.2cm of validators] {};
\node[below=-2.2 of validators] {\bf\dots \qquad \dots};
\node[validator,below=-1.5cm of validators] {};
\node[validator,below left=-1.5cm and -1.2cm of validators] {};
\node[validator,below right=-1.5cm and -1.2cm of validators] {};

\node (a) [right=2.3cm of client] {};
\node (b) [right=4.1cm of client] {};
\path [->, >=stealth'] (client) edge["\circled{1} Claim"] (a);
\draw [dotted, thick] (a.west) -- (b.west);
\path [->, >=stealth'] (b.west) edge[""] ++(2.8cm,0);

%\node (c) [right=2.3cm of verifiers] {};
%\path [->, >=stealth'] (verifiers) edge["\circled{2} Verify"] (c);
\node (c1) [below right=1.2cm and 0cm of client] {};
\node (c2) [right=2.1cm of c1] {};
\path [->, >=stealth'] (c1.west) edge["\circled{2} Verify"] (c2);
\draw [dotted, thick] (c2.west) -- (b.west);

\node (d) [below=2.8cm of b] {};
\node (e) [below=.75cm of d] {};
\path [<-, >=stealth'] (d.west) edge["\circled{4} Sign"] ++(2.8cm,0);
\path [->,>=stealth'] (e.west) edge["\circled{5} Confirm"] ++(2.8cm,0);

\node (f) [above right=-3.5cm and -.2cm of validators] {};
%\draw[->, loop right, min distance=1.2cm, >=stealth'] (f) edge node{\circled{3} Validate} (f);
\draw[->, thick, >=stealth'] (f) 
  to[out=30, in=-30, loop, min distance=1.5cm] (f);
\node at (f) [right=1.2cm] {\circled{3} Validate} ;

\node (g) [below=1.9cm of f] {};
%\draw[->, loop right, min distance=1.2cm, >=stealth'] (g) edge node{\circled{6} Presettle} (g);
\draw[->, thick, >=stealth'](g) 
  to[out=30, in=-30, loop, min distance=1.5cm] (g);
\node at (g) [right=1.2cm] {\circled{6} Presettle} ;

\node (h) [below=.75cm of g] {};
%\draw[->, dashed, loop right, min distance=1.2cm, looseness=3.2, >=stealth'] (h) edge (h);
\draw[->, thick, dashed, >=stealth'](h) 
  to[out=30, in=-30, loop, min distance=1.5cm] (h);
\node at (h) [right=1.2cm] {\circled{7} Settle} ;

\end{tikzpicture}
\end{center}

%\vspace*{2ex}

\small
Proxy is a helper and only affects liveness, not safety.
It can be a client (the sender of the original message containing the claim sequence, or a different client such as a beneficiary or a verifier or a service provider), a validator, or any combination of these for all or for any of the two services it provides: broadcasting messages to validators and signature aggregation ($\#\{\textit{sig}_1,\textit{sig}_2,...\}$).

\vspace*{2ex}

\circled{1} \circled{2}
Client $a$ claims $c_1\,c_2\,...\,c_k$ by signing message $m\coloneq\langle c_1\,c_2\,...\,c_k,n\rangle_a$, where $n$ is its local nonce (incremented after each message).
An appropriate set of verifiers $T:=\{t_1,t_2,...,t_l\}$ approve $m$ by signing it: $\langle m \rangle_{t_1}$, $\langle m \rangle_{t_2}$, ..., $\langle m \rangle_{t_l}$.
Proxy aggregates the verifiers' signatures and broadcasts $\langle m \rangle_{\#T}$ to all validators.

\vspace*{2ex}

\circled{3}
Upon receiving $\langle m \rangle_{\#T}$ with $m=\langle c_1\,c_2\,...\,c_k,n\rangle_a$, validator $v$ validates $m$ by checking:
signature of $a$ is valid;
$v(a).\textit{nonce} = n$;
$v(a).\textit{pending} \subseteq \{m\}$;
the verifiers' aggregated signature $\#T$ is valid and reaches the client-specified quorum;
and finally, %$c_1\,c_2\,...\,c_k$ is valid in $v.\textit{state}$, that is,
$\defined{c_1\,c_2\,...\,c_k}{v.\textit{state}}$.

\vspace*{2ex}

\circled{4}
Validator $v$ signs $m$, sends $\langle m \rangle_v$ to proxy, and sets $v(a).\textit{pending} \coloneq \{m\}$.

\vspace*{2ex}

\circled{5}
Once a validator quorum is reached on message $m$, proxy creates certificate $\langle m \rangle_\#$ and broadcasts it to all validators.

\vspace*{2ex}

\circled{6}
Upon receiving $\langle m \rangle_\#$, each validator $v$: checks the quorum $\#$; checks $m\not\in v.\textit{presettled} \cup v.\textit{settled}$; adds $m$ to $v.\textit{presettled}$.
Note that this $m$ in Step 6 can be different from the $m$ in Step 4, even for the same client $a$.

\vspace*{2ex}

\circled{7}
Settle any message in $v.\textit{presettle}$ that can be settled:
if $m=\langle c_1\,c_2\,...\,c_k,n\rangle_a$ in $v.\textit{presettled}$ with $v(a).\textit{nonce}=n$ and $\defined{c_1\,c_2\,...\,c_k}{v.\textit{state}}$ then $v.\textit{state}\coloneq\denote{c_1\,c_2\,...\,c_k}{v.\textit{state}}$, increment $v(a).\textit{nonce}$, reset $v(a).\textit{pending} \coloneq\emptyset$, move $m$ to $v.\textit{settled}$.
This step 7 can and should be executed continuously.
\end{minipage}
}

\caption{\label{fig:fastset}FastSet protocol.}
\end{figure}

The validator's goal is to consistently replicate the global state, while at the same time offer services through an API.
Each validator $v$ is assumed to maintain the following data at a minimum, and to offer it through its API at request:
\begin{itemize}
    \item A state $v.\textit{state}$, which is its version of the global state that is replicated in each validator.
    In implementations of FastSet the state may be partitioned in objects and storage locations owned by or associated to different clients, as well as shared objects and locations, but this distinction is not necessary in the presentation of the core protocol, because the weak independence abstraction between claims by different clients suffices.
    Each validator may have a different state at any given moment due to message delays and to the massive parallelism allowed by FastSet.
    Yet, we will show that the various state replicas are consistent and they eventually converge.
    \item Sets $v.\textit{settled}$ and $v.\textit{presettled}$ of client submitted messages, which collectively capture all the knowledge that $v$ has about the claims that have achieved validator quorum in the network.
    The claims in (the messages in) $v.\textit{settled}$ have already been processed by $v$, meaning that their effect on $v.\textit{state}$ has already been applied and $v$ is ready to process the next block of claims by the same client.
    The claims in (the messages in) $v.\textit{presettled}$ have not been processed yet by $v$, but will be processed as $v$ is ready for them.
    We will show that all messages in $v.\textit{presettled}$ will be eventually processed and thus moved to $v.\textit{settled}$.
    \item For each client/account/address $a$ some data $v(a)$ including:
    $v(a).\textit{nonce}$, a sequence number counting the claim settlement requests by $a$;
    $v(a).\textit{pending}$, a temporary placeholder containing at most one (latest valid) message received from $a$.
    FastSet implementations may hold more client-specific data.
    Immediate candidates can be $v(a).\textit{verifiers}$, a set of verifiers that $a$ uses to verify its claim blocks, especially when it uses the same verifiers for all messages, and related, $v(a).\textit{quorum}$, a number of verifiers that $a$ considers sufficient in order for its claim blocks to be considered verified.
\end{itemize}

Figure~\ref{fig:fastset} shows how FastSet works.
It is a seven-step protocol.

\circled{1}
Client $a$ broadcasts a block of claims $c_1\,c_2\,...\,c_k$, paired with its local nonce $n$, in a message $\langle c_1\,c_2\,...\,c_k,n\rangle_a$ that it signs, say $m$.
The nonce helps the validators synchronize with the client, making sure that they execute the client's claims in the intended order.
An honest client should never submit two different claim blocks with the same nonce, because that is the FastSet equivalent of a double spending attempt.
Regardless, the protocol will ensure that the client will only be able to obtain validator quorum on at most one claim block per nonce.
A client submitting multiple claim blocks with the same nonce risks to get stuck in a pending status with the validators and never achieve quorum to make progress.
We do not treat this situation here, but implementations of FastSet may choose to allow clients to clear their pending status for a fee, to give them an exit situation in case their programs have errors but at the same time to discourage them from attempting to break or slow down the protocol.

\circled{2}
Validators will not attempt to validate any received claim block before it accumulates an appropriate number of verifiers which approve it.
This information being public, verifiers and clients have multiple ways to synchronize to get this task done.
The simplest, but also the least efficient and riskiest for the client, is for the client to just submit its message and assume that its verifiers listen to some validator and will thus get informed when the client message was settled, then verify it, sign it and send it to the proxy which then aggregates all their signatures and sends them to the validators.
A more efficient approach is for the client to inform its verifiers as soon as it sends the original message, so they can start their verification process immediately.
An even more efficient approach in some applications could be for the client to send its verifiers its intent to execute a program, also known to the verifiers, so that they can start their verification while the client is still working on generating its claim block, allowing program execution and its verification to take place in parallel.

\circled{3}
Once the validator $v$ has been informed that a client's claim block $c_1\,c_2\,...\,c_k$ accumulated the quorum of verifiers, it can proceed to validating it.
The obvious and cheap checks are the signatures, the nonce, and the verifier quorum.
We have deliberately left open the decision on where the verifiers and the verifier quorum are stored:
some implementations of FastSet may store them in the validator, e.g., $v(a).\textit{verifiers}$ and $v(a).\textit{quorum}$, in which case the check becomes
$|T \cap v(a).\textit{verifiers}|\geq v(a).\textit{quorum}$; others may store them as additional items in the message, besides the claim block and nonce; others may include a special claim in the block, say at the end, claiming the verifiers and quorum.
In all cases, however, the verifiers and their quorum are part of the contract that the application makes with its users, so it is important that the validators enforce them.
The validator also checks that there is no different message sent by the same client with the same nonce that is pending, that is, it checks $v(a).\textit{pending} \subseteq \{m\}$.
In other words, if there is any message pending then it must be the exact same message $m$.
This ensures that the client does not, intentionally or not, attempt to initiate double-spending attacks or non-deterministic computations.
Note that messages from validators can be delayed or lost, so the client or its proxy may attempt to submit $m$ multiple times until they receive validator confirmation, which is why $v(a).\textit{pending}$ is allowed to already contain $m$.
Finally, validator $v$ is ready to check the validity of the actual claim block $c_1\,c_2\,...\,c_k$ in its state $v.\textit{state}$, that is,
$\defined{c_1\,c_2\,...\,c_k}{v.\textit{state}}$.
Note that the validator does not update its state at this stage.
Indeed, the block has not achieved validator (not verifier) quorum yet, which means that the client may still possibly achieve quorum on a different block with other validators.

\circled{4}
Validator $v$ now approves the claim block by signing the original client's message and sending it back.
The validator also sets $v(a).\textit{pending}$ to $\{m\}$, so from here on it will accept no other messages from $a$ until $m$ is settled.
If the validator received the same message $m$ from $a$ that it approved already, it sends its approval again to account for previous messages having been potentially lost.

\circled{5}
When the proxy accumulates validator quorum on message $m$, it creates a certificate $\langle m \rangle_\#$ by aggregating all the received validator confirmations and broadcasts it back to all validators.
This certificate is proof that quorum has been achieved and at this moment the state updates in all the validators are imminent, i.e., cannot be stopped anymore once they receive the certificate.

\circled{6}
When a validator $v$ receives such a valid certificate $\langle m \rangle_\#$, it knows with certainty that a quorum of validators have already validated $m$, so it is safe to process and settle $m$.
But that needs to be done only once, to avoid applying the same updates twice or more, so $v$ first checks that $m$ is not already settled, or considered to be settled, i.e., presettled.
If that is the case, then $v$ should settle $m$.
However, $v$ may not be ready to settle $m$, in the sense that the nonce or the claim block of $m$ may not be valid for $v$.
In fact, there is no guarantee that $v$ even participated in the quorum of $\langle m \rangle_\#$, or that $v$ is even aware of the existence of the client that originally sent $m$, because that particular client and $v$ could have been disconnected by now.
%Or even if $v$ participated in the quorum of $\langle m \rangle_\#$, the sender of $m$ and $v$ could have been disconnected for a period and thus $v$ is behind with its state updates and needs to catch up.
Taking into account all these situations, $v$ only adds $m$ to $v.\textit{presettle}$ for now.

\circled{7}
The liveness result, Theorem~\ref{thm:liveness} (see also 4 in Theorem~\ref{thm:allinone}), tells that if other validators settled more claims than $v$, then $v$ should also be able to make progress and settle more claims.
That is, if $v$ continuously scans $v.\textit{presettle}$ it will eventually find messages which are valid and thus can be settled, because FastSet assumes that all messages are eventually delivered.
Implementations of FastSet will likely include mechanisms for validators to request updates of missing settled claims from other validators and populate their presettled set with them in order to make faster progress.
Note that right before $m$ is settled in this step 7 of the protocol, $v(a).\textit{pending}$ can be either empty or $\{m\}$.
The former happens when $v$ did not get a chance to participate in the quorum on $m$, so $v$ never checked $\defined{c_1\,c_2\,...\,c_k}{v.\textit{state}}$ (e.g., if $a$ had enough balance to make a payment); however, since $v(a).\textit{nonce}=n$, $v$ is ready to settle $m$ as soon as $\defined{c_1\,c_2\,...\,c_k}{v.\textit{state}}$, and it does exactly that.
The latter happens when $v$ has already signed $m$ and was waiting for a quorum including other validators as well to be formed.
The monotonicity result, Theorem~\ref{thm:monotonicity} (see also 3 in Theorem~\ref{thm:allinone}), tells that it is safe to settle $m$ as is, that is, checking $\defined{c_1\,c_2\,...\,c_k}{v.\textit{state}}$ is theoretically unnecessary.
However, we choose to keep this definedness check.
As seen in Section~\ref{sec:apps}, implementations of FastSet may choose to allow applications to disobey the weak independence requirement.
Keeping the definedness check will guarantee that the validator's state stays well-defined, the damage being contained to possibly locking the non-conforming accounts.

\begin{theorem}
\label{thm:allinone}
Under the assumptions in Section~\ref{sec:assumptions}, FastSet is correct, i.e.:
\begin{enumerate}
    \item \framebox{Security} All certificates generated at Step 5 are consistent, that is, only one per client per nonce.
    Therefore, FastSet prevents double-spending.
    \item \framebox{Determinism} The order in which a validator receives the messages is irrelevant.
    \item \framebox{Monotonicity} Once a client can settle a message, it will continue to be able to settle it regardless of what other clients do.
    \item \framebox{Liveness} Validators do not get stuck: if one makes progress, then all do.
\end{enumerate}
\end{theorem}
\begin{proof}
We only prove the security property here and the Byzantine aspect of monotonicity and liveness.
The other properties are deferred to Section~\ref{sec:correctness}, because they require a more formal setting.

Recall that we assumed $3f+1$ validators, with at most $f$ Byzantine.
Recall also that a quorum is any subset of $2f+1$ validators, so there are sufficient honest validators to fill the quorum for any valid message.
That is, even if the Byzantine validators decide to stay silent and not send their signed messages in Step 4, a quorum will still be reached for any valid message.
Therefore, when proving the monotonicity and liveness properties in Section~\ref{sec:correctness}, we can assume that the validator advanced to Step 6.

The Byzantine validators can behave arbitrarily maliciously, though, and they can do so in combination with arbitrarily many cooperating clients which they can create and control.
Note, however, that no invalid message can ever be signed (in Step 4) by any honest validator.
Since any message quorum includes at least one honest validator and since signature aggregation cannot be forged in Step 5, we conclude that only valid messages are received at Step 6, and thus only valid messages can ever be settled by honest validators in Step 7.
Therefore, as far as each client obeys the protocol, that is, issues \textit{one} sequence of claims, all honest validators will eventually observe the exact same sequence.
But what if a client, or its proxy, conspires with the Byzantine validators and attempts to fork its sequence of claims, that is, to send two different messages with the same nonce?
We show that this is not possible.
Indeed, each of the two different messages with the same nonce need to reach quorum of $2f+1$ validators.
Let's assume there are $b\leq f$ Byzantine validators, that is, $3f+1-b$ honest validators.
Then each quorum has at least $2f+1-b$ honest validators.
Suppose by contradiction that the respective sets of honest validators in the two quorums are disjoint.
Then it means that there are at least $2(2f+1-b) = 4f + 2 - 2b \geq 3f + 2 - b$ honest validators, which is a contradiction.
Consequently, the two quorums must have at least one honest validator in common.
But then this honest validator would never sign two different messages with the same nonce, because of the $v(a).\textit{pending}\subseteq\{m\}$ clause in Step 3.
\end{proof}

\newcommand{\Address}{\textit{Address}}
\newcommand{\Input}{\textit{Input}}
\newcommand{\Claim}{\textit{Claim}}
\newcommand{\State}{\textit{State}}
\newcommand{\Claims}{\textit{Claims}}
\newcommand{\Output}{\textit{Output}}
\newcommand{\Script}{\textit{Script}}

\newcommand{\state}{\textit{state}}
\newcommand{\claims}{\textit{claims}}
\newcommand{\update}{\textit{update}}
\newcommand{\script}{\textit{sc}}

\section{Formalization and Correctness}
\label{sec:formalization-correctness}

Here we prove the correctness of FastSet.
Before we present our results, it is worth discussing some of the core concepts used to capture correctness or consistency in distributed systems.
These concepts do not apply directly to our setting but are closely related, and they inspired our formalization and results.

Strong consistency~\cite{lamport1978time} states that every data item update is propagated to all nodes before any other access (each read returns the latest write value, regardless of the node on which the read is executed).
Strong consistency is often used synonymously with linearizability~\cite{herlihy1990linearizability}, which means that the distributed system behaves like one computer.
This is the commonly used notion of protocol correctness in blockchains, which are metaphorically called ``world computers'' to convey their linearizability property, and is achieved using consensus algorithms like Paxos~\cite{lamport1989paxos,lamport1998paxos} and Raft~\cite{ongaro2014raft} among many recent ones.
Strong consistency guarantees both safety and liveness, and it is the most comforting property that a distributed system can have.
Unfortunately, in practice it comes at the expense of performance and scalability, and, in the context of blockchains, it also results in high fees per transaction.
In fact, the single-owner object in \cite{Guerraoui2019,guerraoui2022consensus} and then FastPay~\cite{Baudet2019FastPay} were proposed as alternatives to strongly consistent protocols precisely for these reasons, demonstrating that strong consistency can be significantly weakened for many applications, including payment systems like Bitcoin~\cite{nakamoto2008bitcoin}.
FastSet falls into the same category; i.e., it is not strongly consistent.

Eventual consistency~\cite{vogels2009eventuallyconsistent}, on the other hand, states that data item updates will propagate through the system and eventually be applied to all nodes, given enough time (each read in each node eventually returns the latest write).
Put differently, if no new updates are made to a particular data item, eventually, all nodes will converge on the same value for that item.
Eventual consistency is therefore a liveness property, and it is considered the minimal notion of consistency that still has practical value.
However, eventual consistency tends to be too weak for many applications, especially interactive and reactive systems meant to execute forever, like our FastSet, because it does not say anything about the safety, or consistency, of the intermediate values of the data items.

Strong eventual consistency (SEC)~\cite{shapiro2011crdt} brings safety to eventual consistency, in the following sense: any two nodes that have received the same (possibly interleaved) updates will be in the same state.
SEC is clearly the desired property for FastSet as well.
In FastSet terminology, it says that if two different validators process the same claim sequences for the same clients, then they end up in the same state regardless of how the claim sequences were interleaved.
That is, from the collective perspective of the clients, validators behave deterministically and will not diverge.
Or put differently, they are in ``consensus''.

However, the concurrency model underlying FastSet differs from classic (replica-based) distributed systems models like those in \cite{shapiro2011crdt}.
In distributed systems models, a number of communicating nodes perform a computation together.
Accesses to shared objects impose certain constraints on the distributed computation, such as all the reads seeing the latest write value, but are also opportunities for nodes to share causality information.
In FastSet, validators do not initiate any computation nor exchange any information with each other, neither directly nor indirectly through clients, and there is no causal dependence shared through updates (nor defined or calculated).
Clients drive the protocol by broadcasting claims that may update the validator's states.
The only means of communication in FastSet, if it can even be called so, is that a client claim reaching a validator may be invalid for that validator, in which case the validator either discards it (in Step~3) or postpones it (in Step~6).
An abstraction similar to our weak independence was introduced in \cite{shapiro2011crdt} (Definition 2.6, Commutativity) and used to prove the SEC of their model, but it cannot be directly
used in our formalization.

We prove the following three results for FastSet:
\begin{description}
    \item[Determinism.] The interleaving order in which claims sent by clients are received by validators is irrelevant, provided each claim is valid when received.  That is, the state obtained after their processing is the same.
    \item[Monotonicity.] Once a claim submitted by a client is valid for a validator, it will stay valid no matter what other claims submitted by different clients are processed by that validator.  That is, once valid a claim stays valid.
    \item[Liveness.] Validators will never get stuck due to unfortunate interleaving orderings of claims.  That is, if progress has been made by any other validators for any clients, then the validator in question can also make progress.
\end{description}

These properties, together, guarantee the safety and liveness of FastSet.
It is important, however, to recall two major assumptions underlying our abstract protocol: (1) each client issues a determined sequence of claims with each message submitted for validation, and can only settle one message at a time (nonce); and (2) claims issued by different clients are weakly independent.
FastSet is not concerned with the application-specific semantics or the correctness of the claim block in (1), because the client is expected to clear that with its own verifiers; the FastSet validators take the claim block as simple commands to validate and update their state.
Ensuring weak independence can be arbitrarily complex, depending on the type of claims allowed and the expressiveness of the languages that generate them.
In Section~\ref{sec:apps} we discuss such a language, \SETL, and show through a series of common Web3 example contracts that weak independence can be preserved in practice even if the language does not strictly enforce it.

\subsection{Claims and Claim Sequences}
\label{sec:claims}

\begin{definition}
Let $\State$ be a set of \textbf{states} and $\Claim$ be a set of \textbf{claims} together with a given \textbf{claim semantics} as a denotation function:
$$\denote{\_} \ :\ \Claim \rightarrow [\State \rightharpoonup \State]$$
When $\denote{c}{s} \neq \bot$, written also $\defined c s$, we say that claim $c\in\Claim$ is \textbf{valid} in state $s\in\State$ and in that case $\denote c s$ is the state obtained after \textbf{processing} the claim.
\end{definition}

For example, a claim by Alice making a payment to Bob is valid in a state where Alice has enough funds, and the state obtained after processing this claim updates the balances of Alice and Bob adequately.
Section~\ref{sec:fs-wi} discusses many other examples of claims.
When $\denote c s = \bot$, written also as $\undefined{c}{s}$, we say that state $s$ cannot process claim $c$, or $c$ is {\em invalid} in $s$.
For example, an invalid claim can be some payment claim without sufficient balance, or some mathematical theorem whose proof was not verified by a sufficient number of specialized verifiers, or some computation claim which has an error (say a division by zero), etc.

\begin{definition}\label{def:claim-seq}
Extend semantics and validity to claim sequences as expected: $$\denote\_\ :\ \Claim^* \rightarrow [\State \rightharpoonup \State]$$
where $\denote \epsilon s \coloneq s$ and $\denote {c\,\gamma} s \coloneq \denote \gamma {\denote c s}$.
Extend also the $\downarrow$ notation to sequences of claims as expected: $\defined{\epsilon}{s}$, and $\defined{c\,\gamma}{s}$ iff $\defined{c}{s}$ and $\defined{\gamma}{\denote{c}{s}}$.
Also, $\undefined{\gamma}{s}$ iff not $\defined{\gamma}{s}$.
We use $\leq$ for claim sequence prefix: $\gamma \leq \gamma'$ iff $\gamma$ is a prefix of $\gamma'$, i.e., there is some claim sequence $\tau$ such that $\gamma\,\tau=\gamma'$.
\end{definition}

With the notations in Definition~\ref{def:claim-seq} where $\gamma\leq\gamma'$ is witnessed by $\tau$ with $\gamma\,\tau=\gamma'$, if $\defined{\gamma'}{s}$ then $\defined{\gamma}{s}$, $\defined{\tau}{\denote{\gamma}{s}}$ and $\denote{\gamma'}{s}=\denote{\tau}{\denote{\gamma}{s}}$.

\subsection{Weak Independence}
\label{sec:weak-independence}

We next introduce the key abstraction of FastSet:

\begin{definition}\label{def:weak-indep}
Claims $c,c' \in \Claim$ are \textbf{weakly independent}, written $c \parallel c'$, iff $\forall s\in\State$, if $\defined{c}{s}$ and $\defined{c'}{s}$ then $\defined{c\,c'}{s}$, $\defined{c'\,c}{s}$, and $\denote{c\,c'}{s} = \denote{c'\,c}{s}$.
\end{definition}

Weak independence says that if the two claims are \textit{both} valid in a given state, then their processing is independent of each other: they continue to be valid and the same state is obtained regardless of which is processed first and which second.
Many examples of weakly independent claims were discussed in Section~\ref{sec:fs-wi}, notably asset transfers (e.g., payments).
Note that it could be the case that one of the claims is not immediately valid in some particular state, but becomes valid as soon as the other claim is processed: $\undefined c s$, $\defined{c'}{s}$, $\defined{c'\,c}{s}$.
For example, $c$ may be a payment initiated by Alice with insufficient balance, so it cannot be processed, but $c'$ is a payment from somebody else to Alice, which once settled makes $c$ possible.
This is why we call it ``weak'' independence.

What makes weak independence a powerful construct for replica-based settlement protocols that aim at massive parallelism, like FastSet, is the fact that each time a validator is faced with settling a claim, it already \textit{knows} that the claim is valid, because it either checked its validity locally or has received quorum from other validators that the claim is valid.
This means that each validator can safely settle any claim as soon as it is ready, without worrying about non-determinism or getting stuck.
These crucial properties are proved in Section~\ref{sec:correctness}.

\begin{definition}\label{def:weak-equiv}
Claim sequences $\gamma,\gamma'\in\Claim^*$ are \textbf{weakly equivalent}, written $\gamma \Downarrow\gamma'$, iff $\forall s\in\State$, if $\defined{\gamma}{s}$ and $\defined{\gamma'}{s}$ then $\denote{\gamma}{s} = \denote{\gamma'}{s}$.
\end{definition}

Weak equivalence is reminiscent of the classic notion of ``weak equality'' in partial algebra (see, e.g., \cite{Burmeister1982} for a survey):
two claim sequences are weakly equivalent iff they result in the same state whenever they are \textit{both} valid.
What may be counter-intuitive at first sight is that an invalid claim sequence is weakly equivalent to any other claim sequence, whether valid or not---this meaning of ``weak'' is, however, well-established in the literature.
This implies, in particular, that weak equivalence is \textit{not} transitive: if $\gamma \Downarrow \gamma'$ and $\gamma' \Downarrow \gamma''$ then there is nothing to prevent the existence of a state $s$ such that $\defined{\gamma}{s}$ and $\defined{\gamma''}{s}$, but $\undefined{\gamma'}{s}$ and $\denote{\gamma}{s} \neq \denote{\gamma''}{s}$.
For example, $s$ can be a state where Alice has 11 tokens, $\gamma$ and $\gamma''$ payment claims by which Alice pays Bob 10 tokens and 11 tokens, respectively, and $\gamma'$ some invalid claim (e.g., Alice pays Charlie 100 tokens).

However, weak equivalence has all the other properties of a congruence:

\begin{proposition}
\label{prop:sem-cong}
Weak equivalence is:
\begin{enumerate}
    \item Reflexive: $\gamma \Downarrow \gamma$;
    \item Symmetric: if $\gamma \Downarrow \gamma'$ then if $\gamma' \Downarrow \gamma$;
    \item Compatible: if $\gamma \Downarrow \gamma'$ and $c \in \Claim$ then $c\,\gamma \Downarrow c\,\gamma'$ and $\gamma \, c \Downarrow \gamma' \, c$.
\end{enumerate}
\end{proposition}

\begin{proposition}
\label{prop:sem-comm}
$\forall c,c'\in\Claim$, if $c \parallel c'$ then $c\,c'\Downarrow c'\,c$.
\end{proposition}
\begin{proof}
Let $s \in \State$ such that $\defined{c\,c'}{s}$ and $\defined{c'\,c}{s}$.
Since $\defined{c\,c'}{s}$, it follows that $\defined{c}{s}$.
Similarly, since $\defined{c'\,c}{s}$, it follows that $\defined{c'}{s}$.
Since $c \parallel c'$ and $\defined{c}{s}$ and $\defined{c'}{s}$, it follows that $\denote{c\,c'}{s} = \denote{c'\,c}{s}$.
\end{proof}

The reverse is not true.
For example, if the two claims are the same mutex acquires, then each of them is valid independently, but they are not valid together regardless of the order.
So $c\,c'\Downarrow c'\,c$ vacuously holds, but $c \parallel c'$ does not hold.

The following generalizes Proposition~\ref{prop:sem-comm}:

\begin{proposition}
\label{prop:sem-comm-many}
Let $k\geq 0$ and $c,c_1,c_2,...,c_k\in\Claim$ such that $c \parallel c_i$ for all $1 \leq i \leq k$.  Then
\begin{enumerate}
    \item $\forall s\in\State$, if $\defined{c}{s}$ and $\defined{c_1\,c_2\,...\,c_k}{s}$ then $\defined{c\,c_1\,c_2\,...\,c_k}{s}$, $\defined{c_1\,c_2\,...\,c_k\,c}{s}$, and $\denote{c\,c_1\,c_2\,...\,c_k}{s} = \denote{c_1\,c_2\,...\,c_k\,c}{s}$;
    \item $c\,c_1\,c_2\,...\,c_k \Downarrow c_1\,c_2\,...\,c_k\,c$.
\end{enumerate}
\end{proposition}
\begin{proof}
Recall that $\Downarrow$ is not transitive, so we cannot use Proposition~\ref{prop:sem-comm} repeatedly.

Property 2 follows immediately from 1.
To show 1, let us introduce the well-defined state notations $s_1 \coloneq \denote{c_1}{s}$, $s_2 \coloneq \denote{c_1\,c_2}{s}$, ..., $s_k \coloneq \denote{c_1\,c_2\,...\,c_k}{s}$.

Since $\defined{c_1\,c_2\,...\,c_k}{s}$ it follows that $\defined{c_1}{s}$.
Since $c \parallel c_1$ and $\defined{c}{s}$ and $\defined{c_1}{s}$, it follows that $\defined{c\,c_1}{s}$, $\defined{c_1\,c}{s}$, and $\denote{c\,c_1}{s} = \denote{c_1\,c}{s} = \denote{c}{s_1}$, which implies that $\denote{c\,c_1\,c_2\,...\,c_k}{s} = \denote{c_1\,c\,c_2\,...\,c_k}{s} = \denote{c\,c_2\,...\,c_k}{s_1}$.

Since $\defined{c_2\,...\,c_k}{s_1}$ it follows that $\defined{c_2}{s_1}$.
Since $c \parallel c_2$ and $\defined{c}{s_1}$ and $\defined{c_2}{s_1}$, it follows that $\defined{c\,c_2}{s_1}$, $\defined{c_2\,c}{s_1}$, and $\denote{c\,c_2}{s_1} = \denote{c_2\,c}{s_1} = \denote{c}{s_2}$, which implies that $\denote{c\,c_2\,...\,c_k}{s_1} = \denote{c_2\,c\,c_3\,...\,c_k}{s_1} = \denote{c\,c_3\,...\,c_k}{s_2}$.
Combined with the equality in the previous paragraph, this yields $\denote{c\,c_1\,c_2\,...\,c_k}{s} = \denote{c\,c_3\,...\,c_k}{s_2}$.

Iterating this process, we obtain that $\defined{c\,c_k}{s_{k-1}}$, $\defined{c_k\,c}{s_{k-1}}$, and $\denote{c\,c_k}{s_{k-1}} = \denote{c_k\,c}{s_{k-1}} = \denote{c}{s_k}$, which implies our desired equality $\denote{c\,c_1\,c_2\,...\,c_k}{s} = \denote{c}{s_k} = \denote{c_1\,c_2\,...\,c_k\,c}{s}$.
Since $\defined{c_k\,c}{s_{k-1}}$, i.e., $\defined{c_1\,c_2\,...\,c_k\,c}{s}$, the equality also implies the only missing part, namely $\defined{c\,c_1\,c_2\,...\,c_k}{s}$.
\end{proof}

We next extend weak independence to sets of claims:

\begin{definition}\label{def:weak-indep-sets}
Sets of claims $C,C' \subseteq \Claim$ are \textbf{weakly independent}, written $C \parallel C'$, iff $\forall c\in C$, $\forall c' \in C'$, $c \parallel c'$.
\end{definition}

\subsection{Interleavings}

From here on in the paper, we tacitly assume the following:

\begin{assumption}\label{assumption:addresses}
There exists a set $\Address$, whose elements we call \textbf{addresses} or \textbf{accounts}, which partition the claims into weakly independent partitions:
\begin{itemize}
\item $\Claim \ = \ \bigcup_{a\in\Address} \Claim_a$
\item $\forall a,a'\in\Address$, if $a \neq a'$ then $\Claim_a \cap \Claim_{a'} = \emptyset$ and $\Claim_a \parallel \Claim_{a'}$
\end{itemize}
\end{assumption}

In FastSet, addresses correspond to clients.
In practice, the same client may own several addresses, but that is irrelevant in this paper, so we tacitly identify clients with addresses and take the freedom to interchange the two names for the same concept.
Although some claims are generic in practice, like ``Pythagoras theorem'', for simplicity we assume them tagged with their issuer's address, so that claims issued by different addresses/clients are always different.

\begin{definition}
Given a claim sequence $\gamma \in \Claim^*$ and an address $a\in\Address$, the \textbf{filtering} or \textbf{reduct} of $\gamma$ by $a$, written $\reduct{\gamma}{a}$, is defined inductively by $\reduct{\epsilon}{a} \coloneq \epsilon$, $\reduct{(c\,\gamma)}{a} \coloneq c\,(\reduct{\gamma}{a})$ if $c\in\Claim_a$, and $\reduct{(c\,\gamma)}{a} \coloneq \reduct{\gamma}{a}$ if $c\not\in\Claim_a$.
\end{definition}

\begin{definition}
Let $\gamma,\gamma'\in\Claim^*$ be claim sequences.
We say that $\gamma$ and $\gamma'$ are \textbf{interleaving equivalent}, written $\gamma \equiv \gamma'$, iff $\forall a\in\Address$, $\reduct{\gamma}{a} = \reduct{\gamma'}{a}$.
We say that $\gamma$ is an \textbf{interleaving cut} of $\gamma'$, written $\gamma \sqsubseteq \gamma'$, iff $\forall a\in\Address$, $\reduct{\gamma}{a} \leq \reduct{\gamma'}{a}$.
Finally, if $a\in\Address$ then we say that $\gamma$ is an \textbf{interleaving $a$-maximal cut} of $\gamma'$, written $\gamma \sqsubseteq_a \gamma'$, iff $\gamma\sqsubseteq\gamma'$ and $\reduct{\gamma}{a} = \reduct{\gamma'}{a}$.
\end{definition}

Interleaving equivalence is reminiscent of actor/thread interleaving in concurrent programming, each address acting as an actor/thread executing a sequence of commands/claims.
An interleaving cut $\gamma \sqsubseteq \gamma'$ captures the intuition that $\gamma$ and $\gamma'$ are two attempts to execute determined sequences of claims for each address, where $\gamma$ was cut short or did not advance as much as $\gamma'$.
An interleaving $a$-maximal cut says the cut made maximal progress for address $a$.

\begin{proposition}
\label{prop:syn-comm}
$\forall a \neq a'\in\Address$, $\forall c\in\Claim_a$, $\forall c'\in\Claim_{a'}$,
$c\,c' \equiv c'\,c$.
\end{proposition}

\begin{remark}\label{remark:no-swap}
It is possible that $\gamma \equiv \gamma'$ and in some $s\in\State$, $\defined{\gamma}{s}$ but $\undefined{\gamma'}{s}$.
For example, $\gamma = c_{AB}\, c_{BA}$ and $\gamma' = c_{BA} \, c_{AB}$, where $c_{AB}$ is a payment of 1 coin initiated by Alice to Bob, and $c_{BA}$ is a payment of 1 coin by Bob to Alice.
Then $\defined{\gamma}{s}$ and $\undefined{\gamma'}{s}$ in any state $s$ where Alice has at least 1 coin and Bob has 0.
\end{remark}

Remark~\ref{remark:no-swap} tells us that we cannot arbitrarily interleaving-shuffle independent claims in a valid sequence and expect the resulting sequence to stay valid.
However, as seen shortly in Theorem~\ref{thm:determinism}, if the interleaving-shuffled sequence happens to also be valid then the two sequences yield the same state.

\begin{proposition}
\label{prop:syn-cong}
Interleaving equivalence $\equiv$ is a congruence on $\Claim^*$.
\end{proposition}

\subsection{Correctness}
\label{sec:correctness}

\begin{remark} \label{remark:det}
Let us prepend a payment of 1 coin by Charlie to Bob, $c_{CB}$, to the claim sequences in Remark~\ref{remark:no-swap}, that is, consider $\gamma = c_{CB}\,c_{AB}\, c_{BA} \equiv \gamma' = c_{CB}\,c_{BA} \, c_{AB}$.
Then both $\defined{\gamma}{s}$ and $\defined{\gamma'}{s}$ in any state $s$ where Alice and Charlie own at least 1 coin each.
Moreover, $\denote{\gamma}{s}=\denote{\gamma'}{s}$ (both equal to $\denote{c_{CB}}{s}$).
\end{remark}

The next theorem shows that this property holds in general, that is, no matter how the various claims issued (sequentially) by clients are validly interleaved, the same state is obtained in the end.
That means, in particular, that honest validators are in consensus as soon as they process all the claims, and thus will behave the same way in the future.

\begin{theorem}\label{thm:determinism}
\framebox{Determinism}
$\equiv \displaystyle{\subseteq} \Downarrow$:
$\forall\gamma,\gamma'\in\Claim^*$, if $\gamma \equiv \gamma'$ then $\gamma \Downarrow \gamma'$.
\end{theorem}
\begin{proof}
Induction on $|\gamma|=|\gamma'|$.
The base case when $\gamma=\gamma'=\epsilon$ is trivial.
Suppose $\gamma=c\,\delta$, where $c \in \Claim_a$ for some $a \in \Address$, and $\gamma'=c_1\,c_2\,...\,c_k\,c\,\delta'$ for some $k\geq 0$ and $c_1,c_2,...,c_k \in \Claim \backslash \Claim_a$.
By repeatedly applying Propositions~\ref{prop:syn-comm} and \ref{prop:syn-cong}, we get $\gamma' \equiv c\,c_1\,c_2\,...\,c_k\,\delta'$, which further implies $c\,\delta = \gamma \equiv \gamma' \equiv c\,c_1\,c_2\,...\,c_k\,\delta'$, which further implies that $\delta \equiv c_1\,c_2\,...\,c_k\,\delta'$.
By the induction hypothesis it follows that $\delta \Downarrow c_1\,c_2\,...\,c_k\,\delta'$, and further by Proposition~\ref{prop:sem-cong} that
$\gamma \Downarrow c\,c_1\,c_2\,...\,c_k\,\delta'$.
To show that $\gamma \Downarrow \gamma'$, let $s\in\State$ such that $\defined{\gamma}{s}$ and $\defined{\gamma'}{s}$ and let us show that $\denote{\gamma}{s} = \denote{\gamma'}{s}$.
Since $\defined{\gamma}{s}$ and $\defined{\gamma'}{s}$ imply $\defined{c}{s}$ and $\defined{c_1\,c_2\,...\,c_k}{s}$, by Proposition~\ref{prop:sem-comm-many} we have
$\defined{c\,c_1\,c_2\,...\,c_k}{s}$, $\defined{c_1\,c_2\,...\,c_k\,c}{s}$, and $\denote{c\,c_1\,c_2\,...\,c_k}{s} = \denote{c_1\,c_2\,...\,c_k\,c}{s}$, which imply $\denote{c\,c_1\,c_2\,...\,c_k\,\delta'}{s} = \denote{c_1\,c_2\,...\,c_k\,c\,\delta'}{s} = \denote{\gamma'}{s} \neq \bot$.
Since $\gamma \Downarrow c\,c_1\,c_2\,...\,c_k\,\delta'$ and $\defined{\gamma}{s}$ and $\defined{c\,c_1\,c_2\,...\,c_k\,\delta'}{s}$, we conclude that
$\denote{\gamma}{s} = \denote{c\,c_1\,c_2\,...\,c_k\,\delta'}{s}$ and finally that $\denote{\gamma}{s} = \denote{\gamma'}{s}$.
\end{proof}

Determinism in itself is a must-have property, but it is not sufficient.
For example, what if a client is prevented from settling a claim that is valid now, only because they were not fast enough to submit it when it was checked to be valid and in the meanwhile other clients interacted and changed the state of the system in a way that makes the claim invalid?
For example, suppose that Alice can indeed pay Bob 1 token now, i.e., Alice has at least 1 token balance, but Alice gets delayed for whatever reasons and cannot submit the claim until later.
Alice expects the payment to still proceed, because if any change happened to her balance was that others may have added, but not removed tokens from her balance.
Of course, this assumes Alice makes no other claims in the meanwhile.

\begin{theorem} \label{thm:monotonicity}
    \framebox{Monotonicity}
    $\forall \gamma,\gamma' \in \Claim^*$, $\forall a \in \Address$ such that $\gamma \sqsubseteq_{a}\gamma'$, $\forall s\in\State$ such that $\defined{\gamma}{s}$ and $\defined{\gamma'}{s}$, $\forall c \in \Claim_a$, if $\defined{\gamma\,c}{s}$ then $\defined{\gamma'\,c}{s}$.
\end{theorem}
\begin{proof}
Induction on $|\gamma'|-|\gamma|$.
If $|\gamma'|-|\gamma|=0$ then $\gamma\equiv\gamma'$, so by Theorem~\ref{thm:determinism} $\denote{\gamma}{s}=\denote{\gamma'}{s}$, which means that $\defined{\gamma\,c}{s}$ implies $\defined{\gamma'\,c}{s}$.
Now suppose that $|\gamma'|-|\gamma| > 0$ and let $\gamma'=\delta'\,c'\,\tau$ such that $\delta'$ is the largest prefix of $\gamma'$ with $\delta'\sqsubseteq_{a'}\gamma$ for some $a'\in\Address$.
Such a prefix $\delta'$ indeed exists, precisely because $\gamma \sqsubseteq_{a}\gamma'$ and $|\gamma'|-|\gamma| > 0$, i.e., $\gamma'$ contains at least one $c'\in\Claim_{a'}$ with $|\reduct{(\delta'\,c')}{a'}|=|\reduct{\gamma}{a'}|+1$.
Moreover, we also deduce that $a'\neq a$, and thus $c \parallel c'$, because $|\reduct{(\delta'\,c')}{a}|\leq|\reduct{\gamma'}{a}|=|\reduct{\gamma}{a}|$.
We can now use the induction hypothesis with $\gamma \coloneq \delta'$, $\gamma' \coloneq \gamma$ and $c \coloneq c'$ and obtain $\defined{\gamma\,c'}{s}$.
Since $c \parallel c'$ and $\defined{\gamma\,c}{s}$ and $\defined{\gamma\,c'}{s}$, we deduce that $\defined{\gamma\,c'\,c}{s}$.
Now we can use the induction hypothesis once more with $\gamma\coloneq\gamma\,c'$, noticing that $|\gamma'|-|\gamma\,c'| = |\gamma'|-|\gamma|-1$, and obtain $\defined{\gamma'\,c}{s}$.
\end{proof}

Determinism and monotonicity capture correctness of only one validator at a time: it always converges to the same state and it always allows clients to settle claims that (they know) are valid.
But they do not say anything about the decentralized network as a whole, specifically about whether all validators make progress together.
For example, could it be that some validators get stuck due to unfortunate interleavings?
The next theorem rules that situation out.

\begin{definition}
Claim sequences $\gamma,\gamma'\in\Claim^*$ are \textbf{compatible}, written $\gamma \triangle \gamma'$, iff for any $a\in\Address$, $\reduct{\gamma}{a}\leq\reduct{\gamma'}{a}$ or $\reduct{\gamma'}{a}\leq\reduct{\gamma}{a}$.
\end{definition}

\begin{theorem} \label{thm:liveness}
\framebox{Liveness}
$\forall \gamma,\gamma'\in\Claim^*$ with $\gamma\triangle\gamma'$ and $\gamma'\not\sqsubseteq\gamma$, $\forall s\in\State$ with $\defined{\gamma}{s}$ and $\defined{\gamma'}{s}$, $\exists a\in\Address$, $\exists c\in\Claim_a$ in $\gamma'\backslash\gamma$ such that $\gamma\,c\,\triangle\gamma'$ and $\defined{\gamma\,c}{s}$.
\end{theorem}
\begin{proof}
Scanning $\gamma'$ from left to right, let $c$ be the first claim in $\gamma'$ which does not appear in $\gamma$.
Specifically, let $\delta'\,c \leq \gamma'$ be such that $\delta' \sqsubseteq\gamma$ and $\delta'\,c \not\sqsubseteq\gamma$.
Such a decomposition of $\gamma'$ exists indeed, because $\gamma'\not\sqsubseteq\gamma$ and $\gamma\triangle\gamma'$.
Let $a\in\Address$ be the address of $c$, that is, $c\in\Claim_a$.
Note that $\reduct{\delta'}{a}=\reduct{\gamma}{a}$, because $\delta' \sqsubseteq\gamma$ implies $\reduct{\delta'}{a}\leq\reduct{\gamma}{a}$, and $c$ being the first claim in $\gamma'$ with $\delta'\,c \not\sqsubseteq\gamma$ implies $\reduct{(\delta'\,c)}{a} > \reduct{\gamma}{a}$, that is, $\reduct{\delta'}{a}c > \reduct{\gamma}{a}$.
Therefore, $\delta' \sqsubseteq_a \gamma$.
Since $\defined{\delta'\,c}{s}$, because $\delta'\,c\leq\gamma'$ and $\defined{\gamma'}{s}$, Theorem~\ref{thm:monotonicity} implies $\defined{\gamma\,c}{s}$.
Finally, note that $\reduct{(\gamma\,c)}{a}=\reduct{\gamma}{a}c\leq\reduct{\delta'}{a}c=\reduct{(\delta'\,c)}{a}\leq\reduct{\gamma'}{a}$ and
$\reduct{(\gamma\,c)}{a'}=\reduct{\gamma}{a}$ for all $a'\neq a$, which imply $\gamma\,c\,\triangle\gamma'$.
\end{proof}

\section{Examples and Applications}
\label{sec:apps}

Recall from Section~\ref{sec:fs-wi} that FastSet is defined parametrically on top of the abstract notion of a \emph{claim}, and that processing a claim can potentially cause a \emph{state} change.
A concrete design for claims and states was not necessary for proving the correctness of the core FastSet protocol.
However, to implement FastSet and build applications, we need a claim language and concrete states.

Here we introduce such a candidate language called \SETL in terms of lots of examples. 
Programs in \SETL are expected to be small, for which reason we call them \textit{scripts}.
We warn the reader, however, that some scripts can be non-trivial---due to the massively parallel nature of FastSet and the fact that validators may have different states at the same time.
We show that a series of applications that are relevant for Web3, e.g., multi-sig, voting, escrow, verifiable computing, custom assets and payments, auctions, app-chains and blockchains,
can be implemented as \SETL scripts.
Here we use these as examples to explain, informally, the semantics and intuition behind \SETL.
Appendix~\ref{sec:understanding-SETL} includes explanations that help the reader
to better understand the \SETL language. 

We remind the reader that the most common blockchain functionality, that of a payment system over a native token (e.g., Bitcoin),
is a native support, as explained in Section~\ref{sec:fs-wi}.
\SETL provides a primitive \verb|transfer(to,value)| for this; e.g., if \verb|alice| wants to transfer 3 native tokens to \verb|bob|, then she submits and settles the claim $\langle \verb|transfer(bob,3)|\rangle_\texttt{alice}$.

\subsection{Creating Accounts}
\label{sec:creating-accounts}

Creating accounts should be easy, cheap and fast.
We expect and encourage users and applications to create many accounts,
each specialized for a particular purpose: storing some particular data,
executing some particular program, holding some particular assets, etc.
The most basic approach to account creation is to create a user-driven account by providing a public/private key pair,
where the public key is the account name/number.
The user should transfer some tokens to the newly created account, so they can submit claims from there.

Because FastSet allows accounts to have a set of verifiers, a user can always list their main user-driven account as the verifier of other user-driven accounts they create, with quorum of 1, this way being able to prove or take responsibility for claims made from the newly created account.

For example, suppose Alice has a user-driven account as her main account, denoted \texttt{alice}.
She can create another user-driven account denoted
\texttt{alice'} by providing a public/private key pair.
Then, Alice can submit any claim block $C$
from \texttt{alice'} using the corresponding public/private key, as follows:
$$
\langle C \rangle_\mathtt{alice'}
$$
However, there is no guarantee yet that account
\texttt{alice'} is
indeed controlled by Alice.
To establish a connection between the two accounts, Alice
can list her main account \texttt{alice} as the only verifier
for all blocks issued by \texttt{alice'}, by submitting the following block instead:
$$
m = \langle \verb|verify({alice}, 1); |C \rangle_\mathtt{alice'}
$$
Here, $C$ is the block that Alice wants to send from account \texttt{alice'}, while
$$
C' = \verb|verify({alice}, 1); |C
$$
is the actual block that Alice needs to submit.
The command \verb|verify({alice}, 1)| is conditional,
which holds iff the message
has reached a verifier quorum of size 1, with verifiers from the set $\{\texttt{alice}\}$.
That is, account \texttt{alice} must also sign it:
$$
\langle m \rangle_{\#\{\mathtt{alice}\}}
= \langle \langle \verb|verify({alice}, 1); |C \rangle_\mathtt{alice'}
  \rangle_{\#\{\mathtt{alice}\}}
$$
Upon receival of the above message, FastSet validators
check if the block (i.e., $C'$) is valid in Step 3
of Figure~\ref{fig:fastset}.
Block $C'$ is valid iff $C$ is valid and the message has been signed by \texttt{alice}.
The following messages will not be accepted by validators because they are not properly signed
as specified by \verb|verify|:
\begin{quote}
$
\langle \verb|verify({alice}, 1); |C \rangle_\mathtt{alice'}
\qquad\ \ \ \ \qquad\text{/\!/ no verifier sigs}
$

$
\langle \langle \verb|verify({alice}, 1); |C \rangle_\mathtt{alice'}
  \rangle_{\#\{\mathtt{bob}\}}
  \qquad\text{/\!/ incorrect verifier sigs}
$
\end{quote}
However, the following will be accepted and settled:
\begin{quote}
$
\langle \langle \verb|verify({alice}, 1); |C \rangle_\mathtt{alice'}
  \rangle_{\#\{\mathtt{alice,bob}\}}
$
\end{quote}
Therefore, Alice, as the owner of account \texttt{alice},
can convince everyone else that
all claims submitted from \texttt{alice'}
must be issued from and signed by her and nobody else.
Whoever wants to verify that can simply
download all the settled claims issued by \texttt{alice'}
and verify that all of them have the form
$
\verb|verify({alice}, 1); |C
$.
Of course, Alice must remember to wrap her original claims $C$ with the \verb|verify| command, and others must check that all the settled claims issued by \texttt{alice'} are wrapped properly.
A better approach is to use contract-based accounts to achieve the above \emph{by construction}, such that only properly wrapped claims can ever be generated and issued from the side accounts of Alice (see Section~\ref{sec:side-accounts}).

\subsection{Side Accounts}
\label{sec:side-accounts}

A side account is a contract-based account that
has its own separate account state but can only issue claims that require the contract owner to sign.
Side accounts are the proper way to achieve what
account \texttt{alice'} tries to achieve in the previous section.
Suppose Alice has her main account \texttt{alice}
and wants to create a side account.
She can create and settle the following \emph{contract}
by signing and submitting the following message:
$$
\verb|ALICE_SIDE| =
\langle \verb|contract( do(C) { verify({alice},1); C } )| \rangle_\texttt{alice}
$$
Here, \texttt{contract} is the \SETL command for creating a contract based on the definition given in its argument.
We create a contract that has only one contract block called \verb|do(C)| with a parameter \verb|C|, without contract fields, instance blocks, or instance fields.
The parameter \verb|C| can be instantiated by an actual block
and \verb|do(C)| yields the \verb|verify|-wrapped version of that block,
just like how in Section~\ref{sec:creating-accounts}
claim $C'$ is generated from claim $C$.
What is different is that in Section~\ref{sec:creating-accounts},
the creation of $C'$ was manual by Alice,
while here the expansion of \verb|do(C)| is automatic
by FastSet validators.

There can be many ways to implement \verb|do(C)|.
One possible way is to let \verb|C| be a normal argument of type \verb|string| or \verb|bytes|, which holds the
claims as a piece of \SETL code.
In the block body, we wrap the code \verb|C| with an explicit \verb|apply(C)| construct, which indicates the FastSet validators to execute code \verb|C| and generate claims accordingly.
Another possibility is to treat \verb|do(C)| as a macro that is expanded at runtime given an actual block parameter.
We only focus on the capabilities of \SETL here, not on its possible implementations, because our focus is to demonstrate the practicality of FastSet by concrete examples.
We will discuss the concrete syntax, semantics and implementation of \SETL elsewhere.

Once this contract is created and settled, it gets a unique ID, \verb|ALICE_SIDE|, which we interchangeably refer to as the ID or the name of this contract.

The only thing that can happen to \verb|ALICE_SIDE| is for its owner, i.e., Alice, to call its contract blocks, i.e., \verb|do(C)|, with a proper block argument, say \verb|transfer(bob,1)|.
To do so, Alice should sign the following message:
\begin{quote}
$m = \langle \verb|ALICE_SIDE.do(transfer(bob,1)|
\rangle_\texttt{alice}$
\end{quote}
In the \verb|do|-block,
there is the \verb|verify({alice},1)| command, which holds
iff the message has reached a quorum of size 1 with verifiers from the set $\{\verb|alice|\}$;
that is to say, Alice must sign it.
It would be redundant and unnecessary to ask Alice to sign $m$ again like below:
\begin{quote}
$\langle m \rangle_{\#\{\mathtt{alice}\}}
= \langle \langle \verb|ALICE_SIDE.do(transfer(bob,1)|
\rangle_\texttt{alice} \rangle_{\#\{\mathtt{alice}\}}$
\end{quote}
Instead, let us define the intended semantics of \verb|verify| such that
the signature from the original client who issues the claim block, also counts towards the verifier quorum.
Formally, and using the notations in Figure~\ref{fig:fastset}:
Suppose a set $T$ of verifiers collectively signs a claim block from client $a$:
$\langle m \rangle_{\# T} = \langle \langle c_1 c_2 \ldots c_k, n\rangle_a \rangle_{\# T}$.
Suppose for some $1 \le i \le k$, $c_i$ is $\verb|verify(| \mathit{signers}\verb|,| \mathit{quorum} \verb|)|$.
Then $c_i$ is valid iff $\mathit{size}((T \cup \{a\}) \cap \mathit{signers}) \ge \mathit{quorum}$.
That is, the number of parties who have effectively signed the claim block
(i.e., $T \cup \{a\}$) has formed a quorum against the designated verifiers.

Back to our example.
Upon receival of the message
\begin{quote}
$m = \langle \verb|ALICE_SIDE.do(transfer(bob, 1)|
\rangle_\texttt{alice}$
\end{quote}
the validators check that account \texttt{alice} is indeed the creator of the contract \verb|ALICE_SIDE|.
Then, they instantiate the \verb|do|-block with \verb|transfer(bob,1)|, and obtain the actual block:
\begin{quote}
$\verb|verify({alice},1); transfer(bob,1)|$
\end{quote}
Next, the FastSet validators will process this block as if
it is issued from account \verb|ALICE_SIDE|,
which causes them to verify that
$m$ satisfies \verb|verify({alice},1)|.
Since $m$ is signed by Alice, it satisfies the specified quorum.
FastSet validators will transfer 1 token from \verb|ALICE_SIDE| to \verb|bob|
if the balance is sufficient.

The side account \verb|ALICE_SIDE| ensures that
\begin{itemize}
\item It has a separate account state from Alice's main account \verb|alice|;
\item It can only issue the prescribed \verb|do|-blocks,
which require a verification signature from \verb|alice| via the \verb|verify|-command.
\end{itemize}

Before we wrap up, we point out that for side accounts,
we can actually remove the \verb|verify|-command
from the \verb|ALICE_SIDE| contract,
because only the
contract creator can call the contract blocks.
Thus, only Alice can call the \verb|do|-block
of \verb|ALICE_SIDE|.
Even if Charlie gets Alice to sign the message:
\begin{quote}
$\langle \langle \verb|ALICE_SIDE.do(transfer(bob,1)|
\rangle_\texttt{charlie} \rangle_{\#\{\mathtt{alice}\}}
$
\end{quote}
the validators will not accept it because \verb|charlie| is not the creator of \verb|ALICE_SIDE|.
The above attempt message will be rejected/ignored before the \verb|do|-block gets instantiated.
The simplified side account contract is shown below:
\begin{quote}
$\verb|ALICE_SIDE_SIMPLE|:
\langle \verb|contract( do(C) {C} )| \rangle_\texttt{alice}$
\end{quote}

As we will show in the next section, the \verb|verify|-command
will become critical and essential
for writing multi-sig contracts that require
multiple signatures.

\subsection{Multi-Sigs}
\label{sec:app-multisigs}

A multi-sig account is a contract-based account that has its own separate account state
but can only issue claims that require a quorum of signatures from multiple parties.
Multi-sig accounts are extensions of side accounts.
Below, we present various possible ways of creating multi-sig accounts.

\subsubsection{A Simple Multi-Sig Account}

Let us first consider the simplest setting, where Alice wants to set up an account such that
all claims issued from that account must get one extra signature from any of Bob or Charlie,
in addition to her own signature.
Alice can then create and settle the following contract:
\begin{quote}
$
\verb|ALICE_ABC| = $

\ \ \ \ $\langle \verb|contract( do(C) { verify({alice,bob,charlie},2); C } )| \rangle_\texttt{alice}
$
\end{quote}
Then, the only thing that can happen to \verb|ALICE_ABC| is for its creator, Alice,
to call the \verb|do|-block, which requires two verification signatures from Alice, Bob, and Charlie.
Since Alice will sign the call message anyway, effectively it means to seek
one extra verification signature from Bob or Charlie.
This way, \verb|ALICE_ABC| is a multi-sig account.
All claims issued from \verb|ALICE_ABC| must have been approved by Alice
and one of Bob and Charlie.

If Bob and Charlie want their own similar multi-sig accounts, they can create and settle the same contracts:
\begin{quote}
$
\verb|BOB_ABC| = $

\ \ \ \ $\langle \verb|contract( do(C) { verify({alice,bob,charlie},2); C } )| \rangle_\texttt{bob}
$

\vspace*{2ex}

$
\verb|CHARLIE_ABC| = $

\ \ \ \ $\langle \verb|contract( do(C) { verify({alice,bob,charlie},2); C } )| \rangle_\texttt{charlie}
$
\end{quote}

\subsubsection{Using a Template}
\label{sec:templates}

Note that the three contracts \verb|ALICE_ABC|, \verb|BOB_ABC|, and \verb|CHARLIE_ABC| created above
have exactly the same scripts.
A more natural solution that avoids code duplication
is for each of them to create a contract from the same code template:
\begin{quote}
\verb|MULTISIG_ABC: do(C) { verify({alice,bob,charlie},2); C }|

$\verb|ALICE_ABC| = \langle \verb|contract(MULTISIG_ABC)| \rangle_\texttt{alice}$

$\verb|BOB_ABC| = \langle \verb|contract(MULTISIG_ABC)| \rangle_\texttt{bob}$

$\verb|CHARLIE_ABC| = \langle \verb|contract(MULTISIG_ABC)| \rangle_\texttt{charlie}$
\end{quote}
Here, \verb|MULTISIG_ABC| is the unique ID/name of a template claim, which simply settles a piece of code
that can be used to instantiate the actual contract creation claims.
Anyone can submit and settle a template claim such as \verb|MULTISIG_ABC|, so in the above we do not bother to indicate its creator, who can be Alice, Bob, Charlie, or someone else.
The only thing that matters here is that the template \verb|MULTISIG_ABC| can be publicly referenced
by Alice, Bob, Charlie and anybody else, and they can use the template to create their multi-sig contracts.

The three multi-sig accounts  \verb|ALICE_ABC|, \verb|BOB_ABC|, and \verb|CHARLIE_ABC| created above have identical behaviors as the ones created in the previous sub-section.

\subsubsection{Parametric Templates}
\label{sec:parametric-templates}

Above we saw a template, \verb|MULTISIG_ABC|, that was used to create many multi-sig contracts that essentially share the same scripts.
Here we go one step further and introduce templates with parameters.
Parametric templates can be used to create contracts whose scripts are not exactly identical,
but highly homogeneous, and only differ in certain arguments and parameters.

Consider the following parametric template
\begin{quote}
\verb|MULTISIG[SIGNERS,QUORUM]: do(C) { verify(SIGNERS,QUORUM); C }|
\end{quote}
with two parameters \verb|SIGNERS| and \verb|QUORUM|.
Here, \verb|MULTISIG[SIGNERS,QUORUM]| is the unique ID/name of this parametric template claim, which simply settles an uninterpreted piece of code, or list of symbols, that can be used to instantiate the actual contract creation claims.
Then Alice can create her multi-sig account \verb|ALICE_ABC| by instantiating the above template:
\begin{quote}
$\verb|ALICE_ABC| = \langle \verb|contract(MULTISIG[{alice,bob,charlie},2])| \rangle_\mathtt{alice}$
\end{quote}
This multi-sig contract of Alice has identical behaviors as the multi-sig contracts of Alice created
in the previous sub-sections.

From now on, we will declare a template with $k$ parameters as follows:
\begin{quote}
$\verb|TEMPLATE_NAME[P1,P2,...,Pk]: TEMPLATE_BODY|$
\end{quote}
If a template has no parameters, we skip the bracket ``\verb|[| \ldots \verb|]|'' and write:
\begin{quote}
$\verb|TEMPLATE_NAME: TEMPLATE_BODY|$
\end{quote}

\subsubsection{A Dynamic Multi-Sig Contract}
\label{sec:dynamic_multisig}

A more flexible contract for multi-sigs would allow an authority to establish and dynamically modify the signers and the quorum, while the signers would only sign.
This can be done by first creating the following template:
\begin{verbatim}
DYNAMIC_MULTISIG[SIGNERS, QUORUM]:
    signers : Set;
    quorum  : Int;
    constructor {
      signers := SIGNERS;
      quorum  := QUORUM;
    }
    addSigner(user)    { signers.add(user); }
    removeSigner(user) { signers.remove(user); }
    updateQuorum(q)    { quorum := q; }
    instance do(C)     { verify(signers, quorum); C }
\end{verbatim}
This template is intended to be instantiated with a set of signers and a quorum in order to create a contract, as follows:
\begin{quote}
$\verb|CONTRACT_MULTISIG|
=$

\ \ \ \ \ $
\langle \verb|contract(DYNAMIC_MULTISIG[{alice,bob,charlie},2])| \rangle_\texttt{authority}
$
\end{quote}
The contract above has several fields and blocks.
Template \verb|DYNAMIC_MULTISIG| is a template for contracts that have two contract fields,
\verb|signers| and \verb|quorum|,
which will be initialized with the \verb|constructor|-block when the contract is created, and then modified only by the contract creator
with the \verb|addSigner|, \verb|removeSigner| and \verb|updateQuorum| contract blocks.
The \verb|do| block is now an instance block, indicated by the \verb|instance| modifier,
meaning that it can only be called by \textit{instances} of this contract.
Since the other blocks, the constructor and the fields are contract owned, indicated by the missing \verb|instance| modifier, it means that instances of this contract can only produce \verb|do| blocks.

Here is a scenario:
\begin{quote}
%$\verb|CONTRACT_MULTISIG|
%=
%\langle \verb|contract(DYNAMIC_MULTISIG[{alice,bob,charlie},2])| \rangle_\texttt{authority}
%$
%
$\verb|multisig_alice| = \langle \verb|instance(CONTRACT_MULTISIG)| \rangle_\texttt{alice}$

$\verb|multisig_david| = \langle \verb|instance(CONTRACT_MULTISIG)| \rangle_\texttt{david}$

$\langle \verb|multisig_alice.do(transfer(gov,3); transfer(mom,1))| \rangle_\texttt{alice}$

$\langle \verb|CONTRACT_MULTISIG.removeSigner(alice)| \rangle_\texttt{authority}$

$\langle \verb|multisig_david.do(transfer(alice,2))| \rangle_\texttt{david}$
\end{quote}
In words, the \verb|authority| creates a \verb|CONTRACT_MULTISIG|
contract by instantiating the \verb|DYNAMIC_MULTISIG| template,
where 2 out of \verb|alice|, \verb|bob|, and \verb|charlie| need to sign any \verb|do| block initiated by instances of this contract.
Alice and David want to interact with this contract, so they create instances of it.
Alice submits one \verb|do| block for verification, which only requires one more signature from \verb|bob| or \verb|charlie| (because it already has the signature from \verb|alice|).
The \verb|authority| removes Alice from the list of signers, without modifying the quorum, which means that David's last \verb|do| block requires signatures from both \verb|bob| and \verb|charlie| (in addition to \verb|david|'s own signature).
If we only want to allow \verb|signers| to submit \verb|do| blocks, we can modify the \verb|do| block in the template as follows:
\begin{verbatim}
    instance do(C) {
        instance.owner in signers;
        verify(signers, quorum);
        C
    }
\end{verbatim}

All participants in this contract and its instances have to be careful to make sure that their claims can be settled, and are settled correctly as intended.
For example, if \verb|authority| removes too many signers without modifying the contract \verb|quorum| field, then the instances' \verb|do| blocks will never get validated.
Or more subtly, if an instance issues a \verb|do| block before a contract update by \verb|authority| is settled, then different validators may impose different verification requirements and thus the instance's message may not accomplish FastSet validator quorum.

The above is actually a consequence of a deliberate language design decision: \textit{\SETL does not enforce the weak independence property!}
This is an important decision which we expect other implementations of FastSet to also make, so it is worth discussing.
Recall from Sections~\ref{sec:fs-wi} and \ref{sec:weak-independence} that claims that access a shared location, where one is a write, may not be weakly independent.
This is precisely what happens in the \verb|DYNAMIC_MULTISIG| script above: claims issued by the contract owner \verb|authority| may update/write the set of signers and their quorum, while instances of the contract read those shared values in their \verb|verify|.

This can lead to non-deterministic and even unintended validator behaviors.
Indeed, suppose that \verb|david| (and/or his proxy) conspires with corrupted \verb|alice| to get her signature on a transaction that the \verb|authority| and \verb|charlie| would not approve if they knew \verb|alice| were corrupted.
Then \verb|david| (and/or his proxy) submits the transaction claim when \verb|alice| is still a signer (assume that \verb|bob| signs as well), collects and aggregates validator signatures (Steps 3, 4 and half of 5 of the protocol in Figure~\ref{fig:fastset}), but does not broadcast the quorum back to validators until \textit{after} \verb|authority|'s removal of \verb|alice| is settled.
Then \verb|david| broadcasts the message signed by \verb|alice| and settles it against \verb|authority|'s and \verb|charlie|'s will.

Some may argue that the above is not a downside, but a feature of FastSet.
Indeed, it may very well be the case that \verb|david| (and/or his proxy) was not malicious but was simply disconnected right before broadcasting the validator quorum.
Implementations of FastSet may choose to also include the validator aggregated signature together with the message in $v.\textit{presettled}$ in Step 6 of the protocol, and then recheck the validity of \verb|verify| in Step 7.
We choose to not do it in \SETL, for efficiency reasons, but recommend developers and their auditors to think through such scenarios like above and handle them programmatically as desired.
For example, the script can require instances to also provide the current signers and their quorum, and then check them against the contract's:
\begin{verbatim}
    instance do(C, mysigners, myquorum) {
        mysigners == signers and myquorum == quorum;
        verify(signers, quorum);
        C
    }
\end{verbatim}
This adjustment guarantees that the instance's message is invalid if the \verb|authority| changes the signers or the quorum before the instance's message is settled.
While this prevents (potentially) unintended claims to be settled, unfortunate interleavings of the various concurrent actions can lead to an instance being stuck forever as invalid.
This is further discussed and resolved in Section~\ref{sec:breaking-w.i.}.

\subsection{Kudos}

Suppose that a team wants its members to give each other kudos, which come with a monetary value.
Then every member, say Alice, can create a contract for collecting her kudos as follows:
\begin{quote}
\begin{verbatim}
KUDOS_ALICE_CODE:
    k : Int;
    constructor { k := 0; }
    instance pay(n) { transfer(alice, n); k += n; }
\end{verbatim}

$\verb|KUDOS_ALICE| = \langle \verb|contract(KUDOS_ALICE_CODE)|\rangle_\texttt{alice}$
\end{quote}
Here \verb|KUDOS_ALICE_CODE| is a template
and \verb|KUDOS_ALICE| is the actual contract.

Each member who wants to give Alice kudos would then create an instance of the contract and then issue a \verb|pay| block.
For example, Bob settles an instance that is specifically created for
Bob to give Alice kudos:
$$\verb|kudos_alice_bob| = \langle \verb|instance(KUDOS_ALICE)| \rangle_\texttt{bob}$$
and then submits $\langle \verb|kudos_alice_bob.pay(3)| \rangle_\texttt{bob}$.
As a result, 3 native tokens are transferred from
\verb|kudos_alice_bob| to \verb|alice|. 
Note that all accounts, and not just user-driven accounts, 
can hold native tokens. 
To ensure that the instance \verb|kudos_alice_bob| has sufficient tokens,
\verb|bob| (or anyone else) should first transfer some tokens to it. 

The contract field \verb|k|, which counts the kudos,
is stored in the contract, not in the instances.
However, it is shared among all the instances and can be modified by the
\verb|pay|-block of each instance.

To avoid duplication of code, the team may choose to instead share the same contract code template making use of \texttt{contract.owner}.
Then each team member can create their own contract using the exact same code:
\begin{quote}
\begin{verbatim}
KUDOS_CODE:
    k : Int;
    constructor { k := 0; }
    instance pay(n) { transfer(contract.owner, n); k += n; }
\end{verbatim}

$\verb|KUDOS_ALICE| = \langle \verb|contract(KUDOS_CODE)|\rangle_\texttt{alice}$

$\verb|KUDOS_DAVID| = \langle \verb|contract(KUDOS_CODE)|\rangle_\texttt{david}$
\end{quote}
The semantics of \verb|contract.owner| is clear:
it is the address of the owner/creator of that contract, which is unique for a given contract and all its instances.

Finally, since some team members are shy and may not create their own contracts, the team \texttt{lead} may want to create the contracts for everybody using a parametric code template and then share them with the entire team:
\begin{quote}
\begin{verbatim}
KUDOS_CODE[P]:
    k : Int;
    constructor { k := 0; }
    instance pay(n) { transfer(P, n); k += n; }
\end{verbatim}

$\verb|KUDOS_ALICE| = \langle \verb|contract(KUDOS_CODE[alice])|\rangle_\texttt{lead}$

$\verb|KUDOS_DAVID| = \langle \verb|contract(KUDOS_CODE[david])|\rangle_\texttt{lead}$
\end{quote}

\subsection{Settlement of Data and Records}
\label{sec:settle-data}

One of the most basic operations that FastSet implementations provide is to store or record persistent data in fields that are exclusively owned by one account.
Such fields can have different types and additional nested structure, but here we assume only the most basic functionality and defer any non-trivial data-structures and their programmability to language-specific computational claims which will be discussed in later sub-sections.
We also assume that the data and records to be settled are persistent, so they cannot be
modified or updated. 
Allowing modification of settled data is similar to using FastSet as a distributed database, which is more subtle because it can easily compromise
the weak independence property. See Appendix~\ref{sec:database} for more details. 

Suppose Alice wants to settle her name ``Alice Wonderland'' on FastSet.
She can create the following contract:
\begin{verbatim}
ALICE_INFO_CODE:
    name : String;
    constructor {
        name := "Alice Wonderland";
    }
\end{verbatim}
$$\verb|ALICE_INFO| =
  \langle \verb|contract(ALICE_INFO_CODE)| \rangle_\texttt{alice}$$
Then, anyone can use the unique ID/name \verb|ALICE_INFO| as a reference to
the claim that Alice's name is ``Alice Wonderland''. 

Programming languages, virtual machines and programs in these are also data.
Therefore, they can also be settled using FastSet.
We will show how to settle programming languages and programs in Sections~\ref{sec:pls-vms} and~\ref{sec:settleprograms}, respectively.

\subsection{Settling Programming Languages and VMs}
\label{sec:pls-vms}

Many claims are expected to have a computational nature, such as ``This Python program \texttt{fibonacci} on input 10 evaluates to 55''.
In order for such a claim to be adequately verified, Python itself needs to be agreed upon, that is, settled as a verifiable artifact.
The most rigorous such artifact would nevertheless be a formal semantics, paired with a series of expert security and design audits as well as formal properties proved about it using proof assistants and mechanical theorem provers (such as deterministic execution, type safety, etc.).
While we truly hope that the future Web3 applications and users will demand such higher levels of rigor, the truth is that the current state-of-the-art is far from it and it would be unreasonable to require such standards just now.
Additionally, this would happen outside of FastSet anyway, and it ultimately boils down to the same basic principle of verifiers signing claims they are entitled to sign.

In FastSet, a programming language (PL) or a virtual machine (VM) is a particular claim.
The unique identifier, or the hash, of a PL or VM claim serves as the contract between developers writing and claiming programs in that PL or VM, and the verifiers who verify executions of such programs.
There is no unique way to claim a PL or VM; here are some possibilities:
\begin{verbatim}
PYTHON_CODE:
    lang : String;
    bin : Bytes;
    ...
    constructor {
        lang := "Python 3.13.3";
        bin := bytearray;
        ...
    }
\end{verbatim}
$$\texttt{PYTHON}
= \langle \verb|contract(PYTHON_CODE)| \rangle_{\mathit{Python}}$$
and
\begin{verbatim}
EVM_CODE:
    lang : String;
    spec : String;
    ...
    constructor {
        lang := "EVM Prague";
        spec := "module EVM is ...";
        ...
    }
\end{verbatim}
$$\texttt{EVM}
= \langle \verb|contract(EVM_CODE)| \rangle_{\mathit{EF}}$$
The language name and an implementation, such as a binary or source code in some other language, or a formal specification of the language are expected to be included as fields and initialized appropriately.
Ideally, such PL/VM claims are non-ambiguous and self-contained, so that they serve as the source of truth for what the language is.
Also ideally, the owners of the accounts making such claims are publicly known authorities, such as the inventors or co-inventors of those languages, and they should also set up other authorities as claim verifiers, for example the language standards committee members, with high quorums to be reached.
However, all these attributes are external to FastSet.

\subsection{Settling Programs}
\label{sec:settleprograms}

Once a PL/VM is settled as in Section~\ref{sec:pls-vms} or similarly, a supply and demand market can form around it.
On the one hand, developers can use it to write and settle programs, or smart contracts.
On the other hand, program execution verifiers in such PL/VM can offer their services to users of such programs.

Programs can be settled similarly to languages, as claims containing the program as a field.
The precise PL/VM is important, because the same program can have different meaning in different languages or versions of them.
Again, FastSet does not prescribe any specific ways, but here are some examples:
let
\begin{verbatim}
FIBONACCI_CODE:
    lang_addr : Address; // the address at which Python is settled
    pgm : String;
    ...
    constructor {
        lang_addr := PYTHON;
        pgm := "fibonacci(n)...";
        ...
    }
\end{verbatim}
$$\texttt{FIBONACCI} = \langle \verb|contract(FIBONACCI_CODE)|_{\textit{Joe}}$$
be a Python program settled by Joe,
where \verb|PYTHON| is the unique ID/name
of the claim that settles the specific Python language being used,
and the field \texttt{pgm} holds the actual program code.
Let
\begin{verbatim}
UNISWAP_CODE:
    lang_addr : Address; // the address at which EVM is settled
    pgm : String;
    ...
    constructor {
        lang_addr := EVM;
        pgm := "swap...";
        ...
    }
\end{verbatim}
$$\texttt{UNISWAP} = \langle \verb|contract(UNISWAP_CODE)|_{\textit{Hyden}}$$
where \texttt{EVM} is the unique ID/name
of the claim that settles the specific EVM version used in this program.
Both \texttt{PYTHON} and \texttt{EVM} can be settled as shown in Section~\ref{sec:pls-vms}.
Similar to PL/VM claims, program claims can and should also be set up to have verifiers, e.g., security (or ideally formal verification) auditors.

As we will shortly discuss in Section~\ref{sec:verified-computations}, program execution claims consist of a field holding the program claim, e.g., \texttt{FIBONACCI} or \texttt{UNISWAP} above, as well as the input to the program and the states before and after the execution.
Clients can make multiple such computational claims and are responsible for properly composing them in order to avoid divergence (or double spending).
However, such computational claims need to be verified.
The client only proposes the computational claims, but cannot guarantee their integrity.
In fact, the client may generate the claim using a cheap untrusted device, or even no device at all if the result can be inferred differently or guessed.
Or can use powerful machines and parallel infrastructure that generates hundreds of computational claims per second, like some Layer 2 (L2) blockchains do.

As soon as a PL/VM that attracts developers is settled, the demand for verifiers of computations in that PL/VM should increase.
Fortunately, all the verifiers need to know is the PL/VM itself, so they can offer their services as verifiers as soon as the PL/VM is settled.
What could be appealing for verifiers is the fact that they can verify computations generated by various programs at the same time, to keep their hardware effectively used.
For example, Ethereum has 1.2 million nodes currently, each of them verifying the same EVM blocks of transactions.
It takes less than 100ms to verify a block, yet the block time is 12s, so they are used at under 1\% of their capacity.
Supposing that these 1.2M nodes subscribe as FastSet verifiers for EVM, they would likely be able to service all the demand for EVM program executions across all the blockchains, both L1s and L2s.
Moreover, such an efficient infrastructure could also suggest new, more efficient architectures and implementations of blockchains (see Section~\ref{sec:blockchains}).

\subsection{Verifiable Computing}
\label{sec:verified-computations}
\label{sec:verifiable-computing}

Despite its very limited direct programmability, FastSet makes it easy for clients to bring their own programming languages, virtual machines, and any computational models.
All such languages do, from FastSet's perspective, is to generate claims that need to be verified.
It is the client's responsibility to set their contracts up appropriately, to establish conventions for what languages are, who are the allowed verifiers, and what counts as a proof for them.
We here show some possible ways to do it; sophisticated clients will likely find more creative solutions.

The following template is parametric in a programming language \verb|LANG|, a set of verifiers \verb|VERIFIERS|, and a quorum of verifiers \verb|QUORUM|.
\begin{verbatim}
VERIFIED_COMPUTE[LANG, VERIFIERS, QUORUM]:
    instance compute(pgm, input, s, claims, s', proof) {
        verify(VERIFIERS, QUORUM);
        claim("compute", LANG, pgm, input, s, claims, s', proof);
    }
\end{verbatim}
A language owner (such as the standard committee of a programming language)
can use this template to create a contract for the language
and specify a set of verifiers to verify the computational claims from various applications of the language.
Applications can create instances of the language contract and generate computational claims
using the \verb|compute|-block.
The \verb|compute|-block has six arguments, which we will explain shortly after,
but all it does is to simply
settle these arguments as static data on FastSet, using the \SETL command \verb|claim|,
assuming that a verifier quorum is reached.
The \verb|claim| command simply builds a claim as structured data consisting of its arguments, and settles it.
Here, we set the first argument to \verb|"compute"| to indicate that these claims are intended to be computation claims, but to FastSet, all arguments are meaningless.
They are just static data that should be structured and put in a claim, and then settled.

As an example,
suppose that $\verb|EVM| = \langle \verb|...| \rangle_\textit{EF}$ was settled as a language by the Ethereum Foundation, as in Section~\ref{sec:pls-vms}.
In addition, Ethereum Foundation can bring a large number of verifiers to the market for all applications to use, by creating the following contract:
\begin{quote}
$\verb|EVM_VERIFIERS: {V1,V2,...,V1000000}|$

$\verb|VERIFIED_EVM| = $

\ \ \ \ \ \  $\langle \verb|contract(VERIFIED_COMPUTE[EVM, EVM_VERIFIERS, 1000])| \rangle_\textit{EF}$
\end{quote}
We are aware of the abuse of syntax in the above when we create
the contract \verb|VERIFIED_EVM|, where
the parameter \verb|EVM_VERIFIERS| should be used not as a value,
but as an address.
One must dereference \verb|EVM_VERIFIERS| and create the contract
using the actual value \verb|{V1,V2,...,V100000}|.
In this design document, we decide to live with this abuse of syntax and treat it in a separate document where we nail down a formal specification of the \SETL language and its implementations.

Applications can now create instances of $\verb|VERIFIED_EVM|$ and generate computational claims
using the \verb|compute|-block.
By definition and by construction, all computational claims are built as follows:
$$
\verb|claim("compute", ...)|
$$
but this is merely a convention that all the participating clients,
especially applications and verifiers,
must agree upon.
%FastSet implementations merely provide the \SETL command \verb|claim| for building these generic claims and do not care about their possible meaning to the relevant clients.

An application, such as a Layer 2 blockchain \texttt{L2chain}, can now create an instance \verb|L2ver| standing for ``L2 verified'':
$$\verb|L2ver| = \langle \verb|instance(VERIFIED_EVM)| \rangle_\texttt{L2chain}$$
and then issue computational claims by signing messages like these:
\begin{quote}
$\langle \verb|L2ver.compute(pgm1, input1, s1, claims1, s2, proof1)| \rangle_\texttt{L2chain}$

$\langle \verb|L2ver.compute(pgm2, input2, s2, claims2, s3, proof2)| \rangle_\texttt{L2chain}$
\end{quote}
Each of these messages must also be signed by a quorum of the verifiers
for FastSet validators to successfully settle them:
$$\langle \langle \verb|L2ver.compute(pgm1, input1, s1, claims1, s2, proof1)| \rangle_\texttt{L2chain}
 \rangle_{\# T_1}$$
$$\langle \langle \verb|L2ver.compute(pgm2, input2, s2, claims2, s3, proof2)| \rangle_\texttt{L2chain}
  \rangle_{\# T_2}$$
Here, $T_1$ and $T_2$ are two (possibly different) sets of 1000 verifiers in \verb|EVM_VERIFIERS|.
The actual claims that FastSet validators will settle upon receiving the above messages would be the following:
$$\langle \verb|claim("compute", EVM, pgm1, input1, s1, claims1, s2, proof1)| \rangle_\texttt{L2chain}$$
$$\langle \verb|claim("compute", EVM, pgm2, input2, s2, claims2, s3, proof2)| \rangle_\texttt{L2chain}$$

Computational claims like above are meaningless for FastSet, but their settlement indicates that the verifiers approved them.
%Indeed, (a quorum of) verifiers have to approve the \verb|L2ver.compute| blocks, whose  are everything for the verifiers, because they are specialized for verifying these blocks and get paid.
Note that the verification of computational claims and calls into the \verb|L2ver.compute| blocks take place in the verifiers, and thus outside FastSet.

We now explain the intuition behind the arguments of the \verb|compute| claims:
\begin{itemize}
\item \verb|"compute"| - a type code that indicates that this is a computational claim;
\item \verb|LANG| - the programming language, such as \verb|EVM|, settled as in Section~\ref{sec:pls-vms};
\item \verb|pgm| - the program/code that is executed, settled as in Section~\ref{sec:settleprograms};
\item \verb|input| - the input of the program;
\item \verb|s| - the initial state of the program;
\item \verb|claims| - the claim block that the app (such as \verb|L2ver|) generates as part of the execution of \verb|pgm|---this block can be empty if no claims are generated;
\item \verb|s'| - the final state of the program;
\item \verb|proof| - a proof of the computation.
\end{itemize}
Note that \verb|s| and \verb|s'| are states of the program, and not of the FastSet validators.
All the arguments except \verb|claims| are self-explanatory.
The \verb|claims| argument corresponds to the claims that the application generates as the intended byproduct of the execution of program \verb|pgm|, besides modifying its state.
In fact, these claims should be considered part of the semantics of the program by the verifiers, and applications will likely provide specialized SDKs to their devs to generate these claims programmatically.
For example, they can include payments (i.e., \verb|transfer| claims) that the application determined to make while executing \verb|pgm|, in particular to its verifiers who might not sign otherwise.

However, the \verb|claims| arguments in the computational claims
are merely static data for now, and they are not yet processed by FastSet validators, and their effects
are yet to take place.
They are simply stored as static, structured data as part of the computational claims,
as the latter are being produced by apps, verified by designated verifiers, and settled by FastSet validators.
These app-claims will be processed later,
when they are glued together into one deterministic sequential computation
using another contract called \verb|FINALIZE_SEQUENTIALLY|, which we explain later.

The reason why we do not want to apply these application generated claims right away is because we want to enable massive parallelization.
Computational claims are expected to be generated by applications at a higher rate than they can be verified by the verifiers.
Also, computational claims are likely to be generated sequentially, while their verification can be done in parallel.
%Therefore, we want to generate computational claims fast and sequentially, and verify and settle them in parallel.
However, we cannot do this with only one \verb|VERIFIED_EVM| instance, because it cannot issue a new claim block before the previous one is completely processed and settled.
In FastSet, claim settlement for one account is always linear.
The account nonce is used for this purpose.
To enable parallel verification and settlement of computational claims, we must use multiple instances, from which verified computational claims can be settled in parallel:
\begin{quote}
$\verb|L2ver1| = \langle \verb|instance(VERIFIED_EVM)| \rangle_\texttt{L2chain}$

$\verb|L2ver2| = \langle \verb|instance(VERIFIED_EVM)| \rangle_\texttt{L2chain}$

$\langle \verb|L2ver1.compute(pgm1, input1, s1, claims1, s2, proof1)| \rangle_\texttt{L2ver1}$

$\langle \verb|L2ver2.compute(pgm2, input2, s2, claims2, s3, proof2)| \rangle_\texttt{L2ver2}$
\end{quote}
If there are sufficient verifiers to choose from and they are incentivised to reach quorum,
an application can always spawn multiple instances to process its tasks in parallel.
In the above, \verb|L2chain| creates two \verb|VERIFIED_EVM| instances and thus doubles the speed of
computational claims being verified and settled.
Computational claims can now be generated sequentially, but submitted for verification and settlement in parallel, by the two separate instances, without waiting for one to be verified and settled, before the other does so.
Once these two computational claims are verified by possibly two different quorum sets of verifiers,
the following claims will be eventually settled on FastSet:
$$\langle \verb|claim("compute", EVM, pgm1, input1, s1, claims1, s2, proof1)| \rangle_\texttt{L2ver1}$$
$$\langle \verb|claim("compute", EVM, pgm2, input2, s2, claims2, s3, proof2)| \rangle_\texttt{L2ver2}$$

In response to high demand for verification services, verifiers can use multiple shards and specialized hardware to verify many claims in parallel and thus increase their throughput.
With careful splitting of large computations and verifier allocation, applications can offer nearly real-time verified computing, where the only noticeable overhead is the latency added by one claim verification.

Now we can answer the previous question about why the application generated \verb|claims| could not be processed when the computational claims were settled.
It was because, in a massively parallel setting, computational claims are issued and verified at different instance accounts, in parallel.
Therefore, these various fragments of computation must now be glued back together into one
deterministic sequential computation.
We can do this by creating a contract
$$\verb|L2SEQ| = \langle \verb|contract(FINALIZE_SEQUENTIALLY[L2genesis])| \rangle_\texttt{L2chain}$$ using the following template,
where \verb|L2genesis| is the genesis state of this L2:
\begin{verbatim}
FINALIZE_SEQUENTIALLY[GENESIS_STATE]:
    state : Bytes;
    constructor {
        state := GENESIS_STATE;  // the genesis state of the app
    }
    step(pgm, input, s, claims, s', proof) {
        state == s;
        isSettled(
          claim("compute", EVM, pgm, input, s, claims, s', proof)
        );
        claims; // application claims are actually processed here
        state := s';
    }
    getState() { claim("currentState", state); }
\end{verbatim}

On initialization, \verb|L2SEQ| is at state \verb|L2genesis|, which is its genesis state.
Then, \verb|L2SEQ| can make one step forward by calling and settling the \verb|step|-block.
The \verb|step|-block  will eventually update the state of \verb|L2SEQ| to a new state,
if there is a computational claim that has been settled (and thus verified) whose ``start state'' \verb|s|
is identical to the current state of \verb|L2SEQ|.
Additionally, the effects of the \verb|claims| are now being processed by FastSet validators,
before the new \verb|state| is updated to \verb|s'|.
Because the \texttt{state} field of the instance is not accessible to other clients who may want to query it, the contract also includes a \verb|getState| block that allows the instance to claim its current state.

Note that it is possible that multiple computational claims are verified and settled on FastSet and all of them share the same initial state, say \verb|s|, but lead to different final states, say \verb|s'|, \verb|s''|, and so on.
This is similar to a forking situation.
It is fine to have forking computational claims because it is indeed the case that all of them are correct, computationally speaking.
However, the application must pick one and only one of these computational claims to make a \verb|step|, from the \verb|L2SEQ| instance.
Once a computational claim is picked, a \verb|step| is made, and a new state, say \verb|s'|, is reached, the other computational claims with the initial state \verb|s| can no longer be applied to step further.

The \verb|isSettled| command checks if a claim has been settled. 
Every FastSet validator maintains a settled claim set. 
New claims, once they are validated and settled, will be added to this set, 
but no claims can be removed from the set.
In this way, the settled claim set forms an CRDT (see Section~\ref{sec:crdt}) w.r.t.
claim settlement. 
In addition, \verb|isSettled| is a side-effect free Boolean check on the 
monotonically increasing settled claim set, so once it is valid, it continues to stay valid. 
It thus guarantees that using \verb|isSettled| to build a conditional claim does not
break the weak independence property. 

We glossed over many details here, such as how the programs and their inputs are represented, whether the states are given in clear or as indexes in an agreed upon data availability (DA) layer, even over what constitutes a proof and how applications generate the claim arguments first place in order to present them to the verifiers.
All these are FastSet-external conventions which are expected to be agreed upon in advance among the participating stakeholders (language committees/standards, applications, execution service providers, provers, verifiers, etc.).

In Section~\ref{sec:blockchains} we build upon the verified computing approach discussed in this section and show how an actual blockchain can be implemented in \SETL.

\subsection{Voting}
\label{sec:app-voting}

Voting is common on blockchains, yet it is an inherently parallel activity in which the order of votes is irrelevant, making FastSet ideal for it.
An authority needs to set up and create a vote poll and then allow voters who are verified to vote.
Finally, the authority should announce the results.

Let us consider a simple voting scenario, where the authority, say \texttt{gov}, who is also responsible for verifying the voters, wants to accumulate \verb|MIN| valid votes to pass some \verb|MOTION|.
Then \texttt{gov} can create a contract using the \SETL script:
\begin{verbatim}
BASIC_VOTE[MIN]:
    voted : Set;  // users who voted
    constructor {
        voted := {};
    }
    instance constructor {
        not(instance.owner in voted); // guard: must be true
        verify(contract.owner, 1);    // authority verifies voter
        voted.add(instance.owner);
    }
    passed() { size(voted) >= MIN; }
\end{verbatim}

$
\texttt{MOTION} = \langle \verb|contract(BASIC_VOTE)| \rangle_\texttt{gov}
$\\

All a voter has to do in order to vote is create an instance of this contract:

\begin{quote}
% $\texttt{poll} = \langle \verb|contract(VOTING)| \rangle_\texttt{gov}$
%$\texttt{POLL} = \langle \verb|contract(VOTING)| \rangle_\texttt{gov}$

$\langle \verb|instance(MOTION)| \rangle_\texttt{alice}$

$\langle \verb|instance(MOTION)| \rangle_\texttt{bob}$

$\langle \verb|instance(MOTION)| \rangle_\texttt{charlie}$

...

%$\langle \verb|poll_alice.register()| \rangle_\texttt{alice}$
%
%$\langle \verb|poll_alice.vote()| \rangle_\texttt{alice}$
\end{quote}
When an instance is created, the \verb|instance.owner| is verified by \verb|gov| and added to the set of \verb|voters|.
No other instance of the same contract can be created by the same voter, that is, each voter can only vote once.

The contract owner can announce the result whenever \verb|MIN| is reached:
\begin{quote}
    $\langle \texttt{MOTION.passed()} \rangle_\texttt{gov}$
\end{quote}

The toy voting script above was overly simplified on purpose, to highlight the concurrent programming capability of FastSet and \SETL.
In practice, voting contracts will likely separate vote registration from voting itself, allow voters to vote for several options, have weighted votes, have a deadline, etc.

We would like to take the opportunity to further discuss this simple voting contract below, because it touches upon a few important aspects of FastSet that make it different from blockchains.

First, unless \texttt{gov} maintains a list of voters they verified off-set, it is impossible to precisely know the number of votes.
Indeed, even if \texttt{alice} reaches validator quorum on $\langle \texttt{instance(MOTION)} \rangle_\texttt{alice}$, she or her proxy may have delayed sending the certificate back to validators, so she might have never been added to the set of \verb|voters| on some validators.
Even if that certificate has been sent, some validators may be delayed in their execution of the Step 7 of FastSet and have not updated their state yet.
The crucial guarantee, however, is that no validator will settle $\langle \texttt{instance(MOTION)} \rangle_\texttt{alice}$ twice.
If voters need to prove that they indeed voted, the contract can be modified by adding a new block
$$\verb|instance voted() { instance.owner in voted; }|$$
and now \texttt{alice} can settle $\langle \verb|motion_alice.voted()| \rangle_\texttt{alice}$ and use its certificate as proof---here \verb|motion_alice| is the instance $\langle \verb|instance(MOTION)| \rangle_\texttt{alice}$.

Similarly and for the same reasons, it is impossible for \verb|alice| to know with certainty whether her vote was actually counted, that is, that it contributed to claim $\langle \texttt{MOTION.passed()} \rangle_\texttt{gov}$ being settled.
This is the case even if \verb|alice| settled a claim $\langle \verb|motion_alice.voted()| \rangle_\texttt{alice}$ like discussed above chronologically before $\langle \texttt{MOTION.passed()} \rangle_\texttt{gov}$ on some validator(s) that she uses to observe the voting process.
Since chronological order is useful in cases like above, and for other reasons, in Section~\ref{sec:timestamping} we propose to extend FastSet and \SETL with claim timestamping: claim issuers timestamp each claim block, and validators check and approve the timestamp.

Finally, it is insightful to note that although it may appear that a (malicious or not) voter can attempt to create two different instances and submit them concurrently to the validators in the hope that they both get settled, that is not possible thanks to the fact that each account has a nonce and at most one message is allowed in a validator's pending for any given account (steps 3 and 4 in Figure~\ref{fig:fastset}).
A voter attempting to create two concurrent (same nonce) instances would therefore risk to get their account stuck.
A voter attempting to create two concurrent (same nonce) instances would therefore risk to get its account stuck.
There is nothing to prevent them to attempt to create two instances in sequence, each issued with a different nonce, but in that case they are processed in order and the second one will fail on all validators as soon as they processed the first one, as intended.
There is nothing to prevent voters from attempting to create two instances in sequence, each issued with a different nonce, but in that case they are processed in order and the second one will fail on all validators as soon as they processed the first one, as intended.

\subsection{Games}
\label{sec:games}

Generally speaking, games,
especially game scores and game data that are verifiably produced
by a set of cooperating participants,
can be stored in a contract, where each participant creates an instance only to claim
their next move.

As an example, here is a contract for playing chess.
The contract is parametric in the two players, \verb|WHITE| and \verb|BLACK|, as well as in an \verb|ARBITER| whose role is to verify the moves.
The players alternate and each makes a move according to a convention agreed upon with the arbiter, e.g., using the algebraic notation~\cite{fide_laws_2023}---which also includes notations for the end of game (winner or deuce).
\begin{verbatim}
CHESS[WHITE, BLACK, ARBITER]:
    game : String;
    turn : String;
    constructor {
        game := "";
        turn := "WHITE";
    }
    instance white(move) {
        turn == "WHITE"; instance.owner == "WHITE";
        game += move; verify(ARBITER,1); turn := "BLACK"; }
    instance black(move) {
        turn == "BLACK"; instance.owner == "BLACK";
        game += move; verify(ARBITER,1); turn := "WHITE"; }
\end{verbatim}
To play a game, \verb|alice| and \verb|bob| pick an arbiter \verb|charlie| (say a service provider accredited by FIDE) and create the following contract
\begin{quote}
$\verb|GAME1723| = \langle\verb|contract(CHESS[alice, bob, charlie])|\rangle_\textit{any}$
\end{quote}
It is irrelevant who creates the contract---can be any of the three participants, or some other application.
Once created, the contract will store the shared \verb|game| and a shared flag \verb|turn| used to enforce player alternation.
To play \verb|GAME1723|, \verb|alice| and \verb|bob| need to create (empty) instances of it, say
\begin{quote}
    $\verb|crashing_bob| = \langle \verb|instance(GAME1723)| \rangle_\texttt{alice}$

    $\verb|alice22| = \langle \verb|instance(GAME1723)| \rangle_\texttt{bob}$
\end{quote}
and now can start playing by settling alternating moves, e.g.,
\begin{quote}
$\langle\verb|crashing_bob.white("1.Nf3 ")|\rangle_\texttt{alice}$

$\langle\verb|alice22.black("Nf6 ")|\rangle_\texttt{bob}$

$\langle\verb|crashing_bob.white("2.c4 ")|\rangle_\texttt{alice}$

$\langle\verb|alice22.black("g6 ")|\rangle_\texttt{bob}$

...

$\langle\verb|crashing_bob.white("41.Kc1 ")|\rangle_\texttt{alice}$

$\langle\verb|alice22.black("Rc2#")|\rangle_\texttt{bob}$ \quad/\!/ \verb|#| means ``check mate''
\end{quote}

It is interesting to observe how this contract respects the weak independence property, and also to reflect on the power that the ``weak'' generalization of independence gives FastSet in general and \SETL in particular.
There are three accounts that issue claims: the contract owner, \verb|WHITE| and \verb|BLACK|.
The contract owner issues only one claim, the first one which creates the contract itself.
None of the other claims could appear before the contract creation claim, so the weak independence property holds when the contract owner is involved.
Now suppose that the contract owner is not involved.
Then the shared \verb|turn| field mutually excludes either \verb|WHITE|'s claim or \verb|BLACK|'s claim, by making either one or the other invalid.
This way, it can never be the case that both \verb|WHITE|'s claim and \verb|BLACK|'s claim are valid, so the weak independence property holds vacuously.

\subsection{Escrow}
\label{sec:escrow}

The prototypical escrow involves three parties: a seller, a buyer, and an agent.
The buyer wants something, say an object, from the seller, which may take some time and effort from the seller.
So the seller does not want to send the object before knowing that the buyer will pay.
The buyer on the other hand does not want to send the funds to the seller, because the seller may not send the object.
Let's assume a simple escrow controlled by an agent who decides where the funds go, but without  direct access to the funds.
The buyer sends the money to the escrow, and the agent pushes one of two buttons: release the funds to the seller (when the seller sent the object), or refund the buyer (when the seller failed to send the object).
The agent needs to create an appropriate instantiation of the following \verb|ESCROW| contract:
\begin{verbatim}
ESCROW[SELLER, BUYER]:
    amount   : Int;
    isFunded : Bool;
    constructor {
        amount := 0;
        isFunded := false;
    }
    instance deposit(v) {
        instance.owner == BUYER;
        not isFunded;
        v > 0;
        transfer(contract,v);
        amount := v;
        isFunded := true;
    }
    release() {
        isFunded;
        transfer(SELLER,amount);
        isFunded := false;
    }
    refund() {
        isFunded;
        transfer(BUYER,amount);
        isFunded := false;
    }
\end{verbatim}
For example, the \verb|agent| creates the \verb|ESCROW_AB| contract for seller \verb|alice| and buyer \verb|bob|.
The buyer has to create an instance of it and then send the funds:
\begin{quote}
$\verb|ESCROW_AB| = \langle \verb|contract(ESCROW[alice,bob])| \rangle_\texttt{agent}$

$\verb|bob_pays_alice| = \langle \verb|instance(ESCROW_AB)| \rangle_\texttt{bob}$

$\langle \verb|bob_pays_alice.deposit(10)| \rangle_\texttt{bob}$
\end{quote}
The funds are now locked in the escrow and the only way to unlock it is for the \verb|agent| to
either \verb|release| it to the seller, or refund it back to the buyer:
\begin{quote}
$\langle \verb|ESCROW_AB.release()| \rangle_\texttt{agent}$ \ \ \ or \ \ \
$\langle \verb|ESCROW_AB.refund()| \rangle_\texttt{agent}$
\end{quote}

Same like in Section~\ref{sec:games}, weak independence holds.
Indeed, there can only be two accounts at a time that issue claim blocks, namely the contract owner (\verb|agent| above) and the seller (\verb|bob| above).
The buyer does not interact with the contract.
The boolean flag \verb|isFunded| mutually excludes either the contract owner's claim or the seller's, by making either one or the other invalid.
This way, it can never be the case that both the contract owner's claim and the seller's claim are valid, so the weak independence property holds vacuously.

\subsection{Non-Native Tokens and Digital Assets}
\label{sec:nonnative_assets}

Below is a simple contract template that mints a token only at contract creation time and the total supply is assigned to the contract owner.
We expect token contracts to be more involved, with flexible minting and verifiers for it (e.g., \texttt{gov}, \texttt{bank}, etc.), possibly with approvals and transfer-from capabilities like in ERC20 tokens.
Our goal here is to keep it simple in order to demonstrate that tokens can be supported by FastSet and how.
\begin{verbatim}
TOKEN[NAME,SUPPLY]:
    name    : String;
    balance : Address -> Int;
    constructor {
        name := NAME;
        balance[contract.owner] := SUPPLY;
    }
    instance transfer_token(to,v) {
        v > 0;
        balance[instance.owner] >= v;
        balance[instance.owner] -= v;
        balance[to] += v;
    };
\end{verbatim}
Here, we name the non-native token transfer function \verb|transfer_token|,
to not be confused with the native transaction function \verb|transfer|.

One can now create a \verb|TOKEN| contract and make a first transfer as follows:
\begin{quote}
$\verb|USDC| = \langle \verb|contract(TOKEN["USDC",1000000])| \rangle_\texttt{circle}$

$\verb|circle_wallet| = \langle \verb|instance(USDC)| \rangle_\texttt{circle}$

$\verb|alice_USDC| = \langle \verb|instance(USDC)| \rangle_\texttt{alice}$

$\langle \verb|circle_wallet.transfer_token(alice,1000)| \rangle_\texttt{circle}$
\end{quote}
Note that the contract owner, \verb|circle|, also needed to create an instance of the contract in order to gain access to claim blocks reserved for instances.
The recipient of the transfer, \verb|alice|, also happened to create an instance wallet, although that were not needed in order to receive transfers.
Indeed, even if \verb|bob| does not have an instance of \verb|USDC|, the following is still possible:
\begin{quote}
$\langle \verb|alice_USDC.transfer_token(bob,100)| \rangle_\texttt{alice}$
\end{quote}
All the non-native tokens and assets are stored in the contract, in its \verb|balance| field.
Whenever \verb|bob| wants to get access to the 100 tokens that \verb|alice| transferred to him, he can simply create an instance:
\begin{quote}
$\verb|bob_USDC| = \langle \verb|instance(USDC)| \rangle_\texttt{bob}$
\end{quote}
and see the 100 tokens in his balance.

A successful token can result in many transfers made by many users in parallel, especially if used for payments.
For example, VISA averages 20,000 transactions per second.
With the advancement of AI and AI agents, it is expected that the demand for higher rates for payments and micro-payments will significantly grow, perhaps to the order of millions of transfers per second.
FastSet is ready for the challenge.
It can settle a theoretically unlimited number of claims per second, including transfer claims like above.
Indeed, there is nothing to prevent two transfers made by different accounts to proceed concurrently---we assume that location updates are atomic, which can be ensured using conventional synchronization mechanisms (e.g., software locks or hardware transactions).

But is the result deterministic, regardless of how the transfers are interleaved and settled by each of the validators?
The theoretical results proved in Section~\ref{sec:formalization-correctness} ensure the correctness of the protocol, including its validator determinism, whenever the weak independence property holds.
Because digital asset transfers in general and payments in particular represent an important use case for FastSet, we prove the weak independence property of the \verb|TOKEN| contract above.
Suppose that two different accounts, \verb|a1| and \verb|a2|, can issue claim blocks
\begin{quote}
$\langle \verb|transfer_token(b1,v1)|\rangle_\texttt{a1}$

$\langle \verb|transfer_token(b2,v2)|\rangle_\texttt{a2}$
\end{quote}
That implies \verb|v1 > 0| and \verb|balance[a1] >= v1|, and respectively,
\verb|v2 > 0| and \verb|balance[a2] >= v2|.
We have to prove that the balances of \verb|a1|, \verb|a2|, \verb|b1|, and \verb|b2| are, respectively, the same no matter in which order the two transfer claims are processed.
We analyze several cases, depending on whether \verb|b1| and \verb|b2| are equal or not, or if any of them or both are equal to any of \verb|a1| or \verb|a2|.
The most common case is that \verb|b1| and \verb|b2| are distinct and also distinct from \verb|a1| and \verb|a2|.
Then in both cases the balances are clearly the same: \verb|a1-v1|, \verb|a2-v2|, \verb|b1+v1|, \verb|b2+v2|.
If \verb|b1==b2==b| and are different from \verb|a1| and \verb|a2|, then the balances are also the same:
\verb|a1-v1|, \verb|a2-v2|, and \verb|b+v1+v2|.
If \verb|b1==a2| and \verb|b2| is distinct, then the balances are also the same:
\verb|a1-v1|, \verb|a2-v2+v1|, and \verb|b2+v2|.
If \verb|b1==a2| and \verb|b2==a1|, then the balances are the same, too:
\verb|a1-v1+v2|, \verb|a2-v2+v1|.
Finally, if \verb|b1==b2==a2| then the balances are:
\verb|a1-v1|, \verb|a2+v1|.
The remaining cases are similar.
Consequently, the \verb|TOKEN| contract above satisfies the weak independence requirement that guarantees its determinism no matter how the transfer claim blocks by different accounts are permuted.
Because the (atomic) addition and subtraction operations on integers are commutative, an even stronger property holds: the individual claims in different blocks can also be further interleaved.
This gives validators the freedom to maximize the parallelism and thus the throughput of token transfers.

\subsection{Sequencing}
\label{sec:sequencing}

Sequencing of actions that may come in arbitrary order is important for a series of applications, like blockchains.
We here show a simple example for how a set of users can register to a sequencing contract, which imposes a total order on them.
How the order is picked is left entirely to the contract owner here, although in practice users of such a contract may require the use of VRFs, timestamps, or other verifiable means.
\begin{verbatim}
SEQUENCING:
    registered : Set;
    sequenced : List;
    constructor {
        registered := empty;
        sequenced := nil;
    }
    instance register() {
        not(instance.owner in registered);
        registered.add(instance.owner);
    }
    pick(a) {
        a in registered;
        registered.sub(a);
        sequenced.next(a);
    }
\end{verbatim}

To register, a user creates an instance of the contract.
Only one registration per user is allowed here, for simplicity, which means that any attempt by a user to register twice will result in the second claim being invalid.
It is important to use \texttt{instance.owner} instead of \texttt{instance} above, because each instance of the contract will have a different \texttt{instance} number.
At any given moment, the contract owner picks a registered user and moves it from the unordered \texttt{registered} to the ordered \texttt{sequenced}.
Like any claim block, \texttt{pick(a)} may be valid for some validators and invalid for others (those which have not received \texttt{a} in their \texttt{registered} yet).
The contract may need to resend the claim block to validators until quorum is achieved.

Note that although two different accounts can read and write a shared location, the contract does not violate the weak independence assumption that guarantees the correctness of the protocol.
The interesting case is when the contract owner issues a claim block \verb|pick(a)| while an \verb|instance.owner| issues a \verb|register()| block.
Suppose that both claims are valid.
Then it must be the case that \verb|a| and \verb|instance.owner| are distinct, because the former is in \verb|registered| while the latter is not.
This means that the order in which the \verb|instance.owner| is added to \verb|registered| and \verb|a| is removed from \verb|registered| is irrelevant.

Orthogonally to weak independence, it is also interesting to analyze the case when \verb|a == instance.owner| above, which can happen when a user \verb|a| was selected by the contract owner, but the user attempts to re-register before \verb|pick(a)| was settled.
The result is that on some validators the user manages to re-register (namely those which have already processed \verb|pick(a)|), while on the others the re-registration is invalid.
The user has the option to re-submit the re-registration claim to the latter validators and will still be able to achieve quorum, i.e., their account is not stuck.
Of course, the user will have to re-pay those validators the settlement fee, to avoid DOS attacks.

\subsection{Time Stamping}
\label{sec:timestamping}

Attaching timestamps to claims would be very useful, both in applications, where timestamps can be used programmatically, and in validators, where the the timestamps can be used to search for specific claims, browse them chronologically, audit them, and so on.
Unfortunately, it is impossible to assign a perfect settlement time stamp to a claim, because each validator receives the claims in different orders and settles them at different times.
One option could be for validators in Step 4 of FastSet (see Figure~\ref{fig:fastset}) to attach their time at signature, and then the proxy to calculate an average or a median of those times at aggregation time.
But in order for this calculation to be trusted, each validator would need to sign a potentially different message including their local timestamp, which significantly increases the complexity and cost of signature aggregation.

An alternative is to have time service providers on the FastSet network, whose job is to verify time claims made by the clients.
Specifically, clients can send messages along with the timestamp they claim for the message and have time verifiers sign the message and verify the claimed timestamp (Step 2 of FastSet in Figure~\ref{fig:fastset}).
Of course, the verifiers would need to allow an error margin to compensate for network delays.\footnote{A national network (WAN) could see delays between 30 and 50 ms, and international connections (WWAN) might have delays between 100 and 300 ms.}
This can be problematic if other verifiers are also needed in the same message.
Variations of FastSet may choose to allow different sets of verifiers for different tasks, each with its own quorum.

Considering that perfect timestamps are off the table and that validators will likely be reasonably good timekeepers anyway, implementations of FastSet may opt for having their validators also validate timestamps---for all claims, not
only those including explicit timestamps.
Specifically, suppose that the messages sent by clients in Step 1 of FastSet have the form $\langle c_1\,c_2\,...\,c_k,n\rangle_a^\tau$, that is, the timestamp $\tau$ is part of the message, just as the nonce $n$, and not a special claim.
Then each validator also checks $\tau$ in Step 3, within an adequate error bound to allow for network delays.
Allowing clients to timestamp their messages has the additional benefit that messages that are late can be discarded at any moment by proxies and validators which may gossip, and, with some care, validators may also clear their $v(a).pending$ when it contains long expired messages.
Finally, this also allows the contract language to have a \texttt{time} construct, which is interpreted by the validators as the time given by $\tau$.

The \texttt{time} construct can be very useful in FastSet contracts.
For example, it can be used to timestamp activities of interest, such as registrations, and then use the timestamp to prove that the activity happened on time, or to even attempt to order the activities chronologically.
Here is an example contract using \texttt{time} to pick the user who registered chronologically first.
In case of a tie on time, the user whose instance ID/name is smaller wins.

\begin{verbatim}
FIRST[START]:
    first : Int;
    time_first : Int;
    instance_first : Int;
    constructor {
        time_first := infinity;
        instance_first := 0;
    }
    instance constructor {
        time > START;
        time < time_first
        or (time == time_first
            and instance < instance_first);
        first := instance.owner;
        instance_first := instance;
        time_first := time;
    }
\end{verbatim}

Here we assumed that claim blocks are executed atomically, that is, different block claims accessing the same shared variable will not interleave their claims.
This is important, for example, in the constructor above where the shared variables \texttt{time} can be read and written by various instances at the same time.
The overall effect on the shared variables will be the same once all instances are settled regardless of the order in which the instances are processed, but it is important that only one instance at a time accesses the shared variables.
We are aware that the atomicity of claim blocks is a very strong assumption and plan as future work to weaken it by adding synchronization objects (mutexes).

\subsection{Auctions}
\label{sec:auctions}

Auctions tend to be complex contracts, where a set of bidders submit their bids for an item.
The auctioneer keeps the highest bid and the highest bidder receives the item, while the other bidders redeem their funds that were out-bidden.
There are many variations of auction contracts, which, to our knowledge, fall in two broad categories when it gets to how the bidders redeem their funds: either they do it themselves, usually by calling a withdraw function when permitted, or the auctioneer sends each of them their bids back when the auction ends.
None of these approaches is possible within \SETL, due to its deliberately restricted nature.
Indeed, there is no way for an account to withdraw any funds from another account, at least not with the current implementations of the native \texttt{transfer} and the contract \verb|transfer_token| block in Section~\ref{sec:nonnative_assets}: the other account has to send the funds explicitly.
Also, \SETL currently has no mechanism to iterate through all the keys of a map (this might be needed, eventually).

We here present a different type of an auction contract, which takes full advantage of the parallel nature of FastSet.
The key insight is that the highest bidder pays the previously highest bidder back, and sends the difference to the contract.
This way, the contract only locks the auctioneer's item and the highest bid, so there is no need to send any funds back to the out-bidden participants.

\begin{verbatim}
AUCTION[ITEM,BIDDING_TIME]:
    stopBiddingTime : Int;
    highestBidder : Address;
    highestBid : Int;

    constructor {
        ITEM.transfer_token(contract, 1);
        stopBiddingTime := time + BIDDING_TIME;
        highestBidder := contract.owner;
        highestBid := 0;
    }

    instance bid(amount) {
        time <= stopBiddingTime;
        if (amount > highestBid) {
            transfer(highestBidder, highestBid);
            transfer(contract, amount - highestBid);
            highestBidder := instance;
            highestBid := amount;
        }
    }

    instance withdraw(amount) {
        transfer(instance.owner, amount);
    }

    end() {
        time > stopBiddingTime;
        ITEM.transfer_token(highestBidder.owner,1);
        transfer(contract.owner, highestBid);
    }
\end{verbatim}

The auctioneer creates the contract above, at the same time sending it the \verb|ITEM| for bidding, as well as a \verb|BIDDING_TIME| for how long bidding is allowed, e.g.,
$\verb|AliceTicket| = \langle \verb|contract(AUCTION[ticket_item,1000])|\rangle_\texttt{alice}$.
For simplicity we assume only one item, which is a token as in Section~\ref{sec:nonnative_assets}.
This item is locked in the contract until the auction ends.
The contract initializes the \verb|highestBidder| as the contract owner, \texttt{alice} in our case, so the contract owner can redeem the item in case nobody bids on it during the specified period.

Bidders create instances of the contract, send funds to their instances in order to bid them, initiate \texttt{bid} blocks through their instances, and finally withdraw from their instances whatever funds were not used for bidding, e.g.,
\begin{quote}
$\verb|auction3| = \langle \verb|instance(AliceTicket)| \rangle_\texttt{bob}$

$\langle \verb|transfer(auction3,100)| \rangle_\texttt{bob}$
--- \verb|bob| sends 100 to its instance

$\langle \verb|auction3.bid(25)| \rangle_\texttt{bob}$
--- \verb|bob| sends 25 to \verb|AliceTicket| (not \texttt{alice})

$\verb|tk_alice| = \langle \verb|instance(AliceTicket)| \rangle_\texttt{charlie}$

$\langle \verb|transfer(tk_alice,50)| \rangle_\texttt{charlie}$
--- \verb|charlie| sends 50 to its instance

$\langle \verb|tk_alice.bid(30)| \rangle_\texttt{charlie}$
--- sends 25 to \verb|auction3| and 5 to \verb|AliceTicket|

$\langle \verb|auction3.bid(40)| \rangle_\texttt{bob}$
--- sends 30 to \verb|tk_alice| and 10 to \verb|AliceTicket|
\end{quote}
Bidding stops after the allowed time, and then the auctioneer (the contract owner) issues an \verb|end()| block which sends the item to the highest bidder and the paid amount from the contract to the contract owner.
For example,
\begin{quote}
$\langle \verb|AliceTicket.end()| \rangle_\texttt{alice}$
--- sends 40 to \verb|alice| and ticket to \verb|bob|
\end{quote}
At any moment during or after the auction ends, the bidders can withdraw any available funds from their instances, e.g.,
\begin{quote}
$\langle \verb|tk_alice.withdraw(50)| \rangle_\texttt{charlie}$
--- \verb|charlie| withdraws its funds

$\langle \verb|auction3.withdraw(60)| \rangle_\texttt{bob}$
--- \verb|bob| withdraws everything left
\end{quote}

The weak independence assumption that guarantees the correctness of FastSet is still obeyed by the contact above, but it is less obvious than in the previous examples.
Before we show that weak independence holds, let us first illustrate the power, but also the complexity of concurrency, in order to appreciate the critical role that weak independence plays.
The potential problem is that the shared field \texttt{highestBid} can be both written and read by different instances.
Consequently, multiple bidders may bid concurrently and, although their individual bids get quorum, they may potentially not be able to settle.
For example, in the scenario above suppose that both \verb|bob| and \verb|charlie| send their first bids at the same time and the validators receive their bid claims at the same time.
Both claims are valid in Step 3 of the FastSet protocol (Figure~\ref{fig:fastset}) regardless of which is processed first, because at that step the validator states are not modified.
All validators therefore sign both claims and quorum is achieved for both.

Suppose now that half of the validators receive/process the two bids in one sequence in their Step 7, say \verb|bob| first and \verb|charlie| second, while the other half in the other sequence, \verb|charlie| first and \verb|bob| second.
In both cases the contract will hold 30 tokens, \verb|charlie| will be the highest bidder, \verb|bob|'s instance balance of 100 tokens is unaffected, and \verb|charlie|'s instance balance is 20 tokens.
However, in the second case, \verb|bob|'s bid never took place, in the sense that the body of the conditional statement was not executed.
But, importantly, \verb|bob|'s bid block was still valid, so in the end all validators are in the same state and with the same claims settled.
Since \verb|bob|'s instance balance was not affected on any of the validators, his next $\langle \verb|auction3.bid(40)| \rangle_\texttt{bob}$ correctly outbids \verb|charile| on all validators.
We leave it as an exercise to the curious reader to notice that weak independence would be violated if we replaced the conditional statement with a guard that invalidated the block when the amount was not higher than the highest bid.
In Section~\ref{sec:breaking-w.i.} we discuss this case in more depth and propose an extension of FastSet that would work with such examples where weak independence is allowed to be temporarily broken.
We also leave it as an exercise to the reader to notice that the weak independence assumption would also be violated if the previously highest bid would be returned directly to the owner of the instance, instead of to the instance itself.
We discuss this case in more depth in Section~\ref{sec:breaking-w.i.} as well.

Let us now prove that the \verb|AUCTION| contract obeys weak independence.
The interesting case is when two independently highest bids are submitted at the same time.
Regardless of the order in which they are processed in a given state, once both are processed the state is the same: the currently highest bidder's instance is paid back, the winner of the two bids becomes the highest bidder, and the state of the loser of the two bids stays unchanged.

\subsection{AppChains and Blockchains}
\label{sec:blockchains}

An appchain or a blockchain is a sequenced list of blocks of transactions,
which are code written in a programming language that modifies
the chain state.
Chain states are usually very large and for that reason
we here choose to only store their hashes in FastSet's validators.
A block is essentially a sequential list of programs, called the transactions of the blockchain,
that changes the current state hash
into the next state hash.
A chain maintains its mempool that holds all the transactions that
have not been added to the chain.
Users can submit their transactions at any time, and these transactions will be added to the mempool.
The chain picks some transactions from the mempool and builds the next block.

We here present a chain contract, \verb|CHAIN[LANG]|, parametric in a programming language \verb|LANG|.
The chain contract below is assumed to work in collaboration with the \verb|VERIFIED_COMPUTE| contract in Section~\ref{sec:verified-computations} that generates verified computational claims about the changes of the chain states.
We also allow the chain to decide when to finish a block, instead of providing a fixed block size.
%This way, the chain can include more transactions in a block when the network traffic is low, and include fewer transactions when the network traffic is packed.

The chain can and will likely have its own users and accounts for them, as well as its own assets and mechanisms to transfer them among its users.
Then conventional bridge smart contracts can be used on the chain to transfer assets from/to other chains or applications which are not necessarily integrated with FastSet (e.g., Bitcoin, centralized exchanges, games, etc).
In other words, in its most simplistic implementation, the chain can use FastSet only as a verifiable settlement layer for its blocks, but not for interacting with other applications or for storing its users' assets.
However, that would be a missed opportunity for the chain, because FastSet offers many, it not most, of the operations that users want to do with their assets, at theoretically uncapped performance.

For that reason and to make the example more interesting, we choose to let the chain interact with FastSet.
Specifically, we allow users to settle FastSet claims together with submitting transactions to the chain, and also allow blocks to settle claims together with updating the chain state.
This opens a broad range of possibilities.
For example, chain users and accounts can hold their assets directly on FastSet as non-native tokens (Section~\ref{sec:nonnative_assets}).
Also, applications can use FastSet for everything that can take advantage of parallelism, and use the chain for everything that requires sequential execution.
For example, an AI agentic application can use FastSet for storing AI agent states, for verifiable communication between agents, and for transferring assets from one agent to another, all at the speed of light, but use a chain (or more) for trading.

%In the following contract, we define the \verb|mempool| to be a map from transaction bytes to \verb|user_claims| represented as strings.
%When the blockchain \verb|pick|s the next transaction, the associated \verb|user_claims| is appended to all the other user claims, all of which will be processed and settled in \verb|add_block()|, when the block is added to the blockchain.

\begin{verbatim}
CHAIN[LANG]
    mempool : Set;
    chain : List;
    next_block : List;

    constructor {
        mempool := {};
        chain := [];
        next_block := [];
    }

    instance submit(trans, user_claims){
        apply(user_claims);  // e.g., bridging assets
        mempool.add(trans);  // assumed in sync with user_claims
    }

    pick(trans) {
        trans in mempool;
        mempool.sub(trans);
        next_block.next(trans);
    }

    add_block(block_claims, next_header, proof) {
        // let VERIFIED_COMPUTE contract generate the next claim
        isSettled(claim("compute",
                    LANG,
                    next_block,    // pgm
                    0,             // input (N/A)
                    chain.last,    // s
                    block_claims,  // claims to process with block
                    next_header,   // s'
                    proof))
        apply(block_claims);
        chain.next(next_header);
        next_block = [];
     }
\end{verbatim}
For concreteness, let us assume the verifiable compute context discussed in Section~\ref{sec:verified-computations}, where \verb|EVM| is settled as a language claim and \verb|VERIFIED_COMPUTE| is a contract that allows its instances to make use of a set of specialized verifiers to verify compute claims using \verb|EVM| and then settle them as claims of the form:
\begin{verbatim}
    claim("compute", EVM, pgm, input, s, claims, s', proof)
\end{verbatim}
where \verb|pgm| is an \verb|EVM| program (sequence of transactions, in our case), \verb|input| is its input, \verb|s| and \verb|s'| are the states before and after the execution of the program, \verb|claims| are the FastSet claims produced by the \verb|EVM| program execution, and \verb|proof| is the proof needed by the verifiers to verify the claim.

To create an \verb|EVM|-based chain using the parametric contract template above, a user or an application \verb|chain| needs to issue a claim as follows:
$$
\verb|evmchain| = \langle \verb|contract(CHAIN[EVM])| \rangle_\texttt{chain}
$$
For simplicity, we here assume that \verb|chain| is a normal FastSet account, which has the freedom to pick any order of transactions.
For increased trust, security and decentralization, in practice this \verb|chain| account may be controlled by a multi-sig, or by an aggregated signature, or at a minimum the \verb|pick(trans)| blocks may require verification from an approved set of qualified verifiers.
We keep the discussion high-level and do not dive into these concrete options here.

To use the blockchain, FastSet users need to create instances of \verb|evmchain|:
\begin{quote}
$\verb|evmchain_alice| = \langle \verb|instance(evmchain)| \rangle_\texttt{alice}
$

%$\verb|evmchain_bob| = \langle \verb|instance(evmchain)| \rangle_\texttt{bob}$
\end{quote}
The instances give them access to \verb|submit| blocks, for example:
\begin{quote}
$\langle \verb|evmchain_alice.submit(alice_trans, transfer(evmchain,1)| \rangle_\texttt{alice}
$
\end{quote}
Above, \verb|alice| submitted a transaction on the chain, \verb|alice_trans|, and 1 token to the contract.
This could be, for example, a bridge transfer, which mints 1 token in \verb|alice|'s account on \verb|evmchain|.
We are not concerned with the integrity or semantics of the specific chain transactions here.
These can be verified by the \verb|chain| contract itself, or by its verifiers, or can be generated to be semantically-correct-by-construction using variants of \verb|submit| blocks, e.g., $\langle \verb|evmchain_alice.bridge(1)| \rangle_\texttt{alice}$ (for brevity, we do not define \verb|bridge| here).

The \verb|mempool| holds the set of pending chain transaction intents, that is, the transactions submitted by its users which were not executed by the chain yet.

\subsection{Beyond Weak Independence}
\label{sec:breaking-w.i.}

In order to be expressive, \SETL does not enforce weak independence.
Instead, it lets contracts prove it, like we did in the previous subsections of this section.
Users should be cautious and avoid getting involved with any contracts that they do not trust or that were not properly audited to prove their weak independence compliance, ideally using formal verification techniques like deductive verification or model checking.
In this section we further discuss the gravity of breaking weak independence and propose an extension to languages like \SETL that could ameliorate the problem in some, but certainly not all, situations.

First, consider the apparently innocent change in the \verb|AUCTION| contract in Section~\ref{sec:auctions}, where the new highest bidder pays back the previously highest bidder \textit{owner}, instead of their instance.
That is, replace \verb|instance| with \verb|instance.owner| in the \verb|bid| block (and remove \verb|.owner| in the \verb|end| block).
Consider the same scenario discussed in Section~\ref{sec:auctions}, where \verb|bob| and \verb|charlie| submit their bids concurrently, and half the validators process \verb|bob| first and then \verb|charlie|, while the other half \verb|charlie| first and \verb|bob| second.
In both cases the contract will hold 30 tokens, \verb|charlie| will be the highest bidder, and \verb|charlie|'s instance balance is 20 tokens.
However, in the first case \verb|bob|'s balance is 25 tokens more while \verb|bob|'s instance balance is 75 tokens, while in the second case 
\verb|bob|'s balance is unaffected while \verb|bob|'s instance balance is 100 tokens.
It may look like there is no problem, because \verb|bob|'s \textit{total} balance in both its account and in its instance stays the same, and he always has the option to withdraw from the instance.
However, \verb|bob| effectively lost 25 tokens, because he will only be able to get validator quorum on withdrawals of up to 75 tokens.
Additionally, from here on, \verb|bob|'s account will show different balances on different validators, without any chance for the validators to ever converge on \verb|bob|'s balance.

Let us next consider another change to the \verb|AUCTION| contract in Section~\ref{sec:auctions}, where the conditional statement in \verb|bid| is replaced with its condition as a guard followed by the same commands.
That is, \verb|bid| is invalid now when the amount is insufficient, instead of silently valid.
Now the case when \verb|bob|'s bid is processed after \verb|charlie|'s is invalid, thus breaking the weak independence assumption.
As a consequence, \verb|bob|'s bid cannot be settled.
In fact, it has no chance to ever be settled for that half of the validators, regardless of how the auction proceeds, not even after the auction ends.
So \verb|bob|'s bid will stay forever in $v(\verb|auction3|).\textit{pending}$ for all those validators.
Not only is this garbage both computationally and space-wise for half the validators, but also it locks the \verb|auction3| account at the current nonce on those validators, which means that \verb|auction3| is effectively forever locked together with \verb|bob|'s 100 tokens, because it will never be able to achieve quorum on any message anymore.

If \verb|bob| still wants to outbid \verb|charlie| he can create another instance of the contract, say
$\verb|auction4| = \langle \verb|instance(AliceTicket)| \rangle_\texttt{bob}$, transfer new funds to it and then
$\langle \verb|auction4.bid(40)| \rangle_\texttt{bob}$.
However, even though in this particular situation \verb|bob| has a solution at the expense of losing the 100 tokens locked in the \verb|auction3| instance, the fundamental problem stays and is clearly rather grave.
Any FastSet implementation that allows accounts to disobey the weak independence assumption, like our \SETL, should probably also provide a general resolution for what to do in situations like the above where the monotonicity property is violated.
Otherwise, malicious attackers can easily initiate DOS attacks on the network by overloading the validators with pending messages that never settle in Step 7 of the protocol.
In \SETL, we choose the following:
\begin{quote}
\textbf{Monotonicity Enforcement Resolution:} In Step 7 of FastSet in Figure~\ref{fig:fastset}, if $m=\langle c_1\,c_2\,...\,c_k,n\rangle_a$ in $v.\textit{presettled}$ with $v(a).\textit{nonce}=n$ and $v(a).\textit{pending}=\{m\}$ but $\undefined{c_1\,c_2\,...\,c_k}{v.\textit{state}}$, then increment $v(a).\textit{nonce}$, reset $v(a).\textit{pending} \coloneq\emptyset$, and move $m$ to $v.\textit{settled}$.
\end{quote}
That is, \SETL tacitly settles the message as if everything was alright, but it ignores its undefined effects on the state.
In particular, the nonce of the account is incremented, this way allowing the account to continue to submit claims---may want to do so to recover.
Note that the account has already paid the settlement fee, so validators do no unpaid work.
In other words, \SETL takes the healthiest decision for itself and leaves it to the apps to semantically fix the problems resulting from using contracts that break weak independence.

\SETL's monotonicity enforcement resolution makes the modified auction contract work as intended: fast, correct, and minimal on resources.
Indeed, each bidder submits their bids unrestricted by anything but the deadline, as it should be, without any sequencing or waiting required, as it should be, and only the contract needs to store data, and only two integers: the highest bidder and its amount.
Correctness means that from all validators' perspectives, the highest bidder gets the item from the auctioneer and all other bidders get their money back.
It follows easily by induction.
Suppose that the property holds for $n$ bidders, and consider one more bid.
Let $v$ be any validator.
If the new bid is higher than the previously highest bid for $v$, then the \texttt{bid} claim block will set the new bid as the highest and the previously highest bidder will be transferred its money back by the new bidder (which can also be the same bidder).
So the inductive property holds in this case.
If the new bid happens to be lower than the previously highest bid for $v$, which can happen because of the receive and settlement order in Step 7 of the FastSet protocol on $v$, then the \texttt{bid} claim block will be invalid and, therefore, by the monotonicity enforcement resolution above, will not change the state of $v$ (but it will increase the nonce of the bidder, so it can continue to participate in bidding).
So the property holds in this case, too.

The second modified auction contract in this section is a good example illustrating why implementations of FastSet, like our \SETL, may choose to not enforce the theoretical weak independence requirement.
While weak independence guarantees the correctness of the protocol in all situations and for all contracts, it may sometimes be too restrictive.
If implementations do not enforce weak independence statically, like our \SETL, then users should be cautious and not use contracts they do not trust or which are not appropriately audited.

\section{Conclusion}
\label{sec:conclusion}

This paper introduced FastSet, a replica-based distributed protocol for embarrassingly parallel settlement of claims.
A \textit{claim} is any statement that comes with a proof, such as a payment or more generally a digital asset transfer, a realized blockchain transaction or more transactions that form a block, an execution of a program in a programming language or virtual machine, an AI model inference or fine tuning, a TEE execution, a vote, an auction bid, among many others.
A \textit{proof} is anything that is verifiable and is accepted by the application or user/account that uses the claim: a signature (simple, aggregated, TEE, etc), a cryptographic/ZK proof, a mathematical proof, a formal semantics derivation, a program (re)execution, and so on.
The verification of all proofs which are not signatures is deferred to special service providers, called \textit{verifiers} which are account holders like any other applications/users.
The replicas, called \textit{validators}, only validate signatures, regarded as the simplest and fastest way to check proofs, and settle the claims.
Each claim can be verified, validated, and \textit{settled optimally}: independently and in parallel with any other claim.
The validators need not communicate with each other, so there is no consensus in the strict sense of the word as used in blockchains.
Specifically, FastSet is not strongly consistent~\cite{lamport1978time}, but
it is strongly eventually consistent~\cite{shapiro2011crdt}.

FastSet drew inspiration from three relatively recent works, published after 2019, which in our view have brought irrefutable evidence that strong consistency in general, and blockchains in particular, are not necessary in order to prevent the double spending attack problem: the consensus number of a cryptocurrency~\cite{guerraoui2022consensus}, FastPay~\cite{Baudet2019FastPay}, and POD~\cite{alpos2025pod}.
These works demonstrated that the peer-to-peer decentralized payment system challenge that was notoriously solved by Bitcoin~\cite{nakamoto2008bitcoin} and has stimulated more than a decade of sustained research, engineering and investments advancing the blockchain technology to what it is today, can also be solved differently.
Quite completely differently, in fact, with no chains (or total order) and no blocks, and without any theoretical limitations on the number of transactions per second.
FastSet brings a general mechanism to settle any claims that are verifiably true, including universal programmability: any computations done using any programming models, languages, or VMs.
At a minimum, FastSet is to FastPay what Ethereum is to Bitcoin.

FastSet is orthogonal to and can be integrated with various execution layers.
These execution layers become FastSet clients/accounts which generate verifiable computational claims,
whose proofs will be broadcast to the relevant verifiers for verification.
Once verified, these computational claims are settled by the validators, and the other clients, such as
L1s, L2s, bridges, games, and so on, can use them to verifiably proceed to their next states.
Fast execution technologies such as MegaETH (\verb|https://megaeth.com|) and Groundhog~\cite{ramseyer2024groundhog}
will be fast computational claims generators on FastSet.
These efficient execution layers also exploit commutativity, as FastSet does,
but in a different setting where transactions must eventually be linearly ordered and processed.

FastSet can be extended or optimized in several directions.
In the current design, FastSet validators use and maintain
their own local state and need not communicate with each other.
Consistency is maintained thanks to weak independence of claims and commutativity of claim settlement.
However, broadcasting to a large number of validators may not be practical in some situations, so
gossip communication among the validators can be allowed
to more efficiently broadcast new claims across the network, using related technologies in coding theory
such as Optimum~\cite{Deb2006AlgebraicGossip}
(\verb|https://getoptimum.xyz|) and set reconciliation~\cite{Yang2024PracticalRateless}.
We currently assume that all validators are involved in all claim validations,
which means that for any claim settlement, a quorum of the entire base of validators must be reached.
Adding more validators to the network, in theory, will not increase the latency of FastSet because validators work independently from each other.
However, it will increase the network traffic.
A possible solution is to adopt a committee-based approach like the one in Algorand~\cite{algorand} 
to enable even more parallel settlement and thus further improve performance.
In the current design, a certified but (temporarily) invalid claim is marked as pre-settled and then queued in Step 6.
In the worst and most unfortunate scenario, the queue can grow indefinitely.
For example, a possibly malicious application may DOS attack some validators by delaying them from seeing messages which would turn
the currently-invalid certified claim valid.
There exist many possible solutions.
A time-based solution relying on the timestamping feature introduced in Section~\ref{sec:timestamping}
may allow validators to discard invalid certified claims that have become outdated and thus allow FastSet to offer a perfect-past consensus result like in POD~\cite{alpos2025pod}.
Another possible solution is to enable the processing and settlement of invalid claims, generalizing
FastPay~\cite{Baudet2019FastPay}, where invalid payments can still be processed and settled, resulting in
(temporary) negative account balances.
We leave these extensions and optimizations as future work.

FastSet should not be regarded as the foundation for the next blockchain.
It should be regarded as the Web3 infrastructure on which the next wave of blockchains and verifiable computing applications will be built.

\paragraph{Acknowledgements.}
The motivation and inspiration for this work came from herculean efforts made by the Pi Squared team to retrofit their (zero-knowledge) Proof of (mathematical) Proof technology to blockchains.
Proof of Proof ($\pi^2$) yields optimal verifiable computing for any programming language, by generating the cryptographic proofs from mathematical proofs produced using an executable formal semantics.  Unlike in the existing zkVM approaches, $\pi^2$ requires no translation or compilation to a specific VM or ISA, and it also allows well-understood mathematical proof engineering to be done before any cryptography.
There was no clean way to bring the full benefits of $\pi^2$ to Web3 through blockchains, because these come hardwired with particular VMs and unnecessarily enforce sequencing, which is often considered subjective and controversial (MEV).
Indeed, $\pi^2$ is truly universal in the PL/VM and sequencing can and should be specified by the apps that need it (see Section~\ref{sec:sequencing}), and not enforced on the entire universe as an artifact of an old belief that it is required in order to avoid double-spending.
Our warmest thanks to the Pi Squared team, whose efforts to avoid the blockchain limitations above have lead to the design of FastSet as a new Web3 infra.
Specifically, we'd like to thank Brandon Moore, Musab Alturki, Traian-Florin Serbanuta, Virgil Serbanuta, Dorel Lucanu, Iulia Bastys, and Ovidiu Damian for suggestions on how to improve this paper.

\bibliographystyle{plain}
\bibliography{refs}

\begin{thebibliography}{10}

\bibitem{Agha1986}
Gul~A. Agha.
\newblock {\em Actors: A Model of Concurrent Computation in Distributed
  Systems}.
\newblock MIT Press, Cambridge, MA, USA, 1986.

\bibitem{alpos2025pod}
Orestis Alpos, Bernardo David, and Dionysis Zindros.
\newblock Pod: An optimal-latency, censorship-free, and accountable generalized
  consensus layer.
\newblock {\em CoRR}, abs/2501.14931, 2025.

\bibitem{Baudet2019FastPay}
Mathieu Baudet, George Danezis, and Alberto Sonnino.
\newblock {FastPay}: High-performance byzantine fault tolerant settlement.
\newblock In {\em {AFT} '20: 2nd {ACM} Conference on Advances in Financial
  Technologies, New York, NY, USA, October 21-23, 2020}, pages 163--177. {ACM},
  2020.

\bibitem{Burmeister1982}
Peter Burmeister.
\newblock Partial algebras---survey of a unifying approach towards a two-valued
  model theory for partial algebras.
\newblock {\em Algebra Universalis}, 15(1):306--358, 1982.

\bibitem{Deb2006AlgebraicGossip}
Supratim Deb, Muriel M{\'{e}}dard, and Clifford Choute.
\newblock Algebraic gossip: a network coding approach to optimal multiple rumor
  mongering.
\newblock {\em {IEEE} Transactions on Information Theory}, 52(6):2486--2507,
  2006.

\bibitem{fide_laws_2023}
{FIDE}.
\newblock Fide laws of chess, 2023.
\newblock Accessed: 2025-06-02. Section E: Algebraic Notation outlines the
  standard for recording chess moves in international competitions.

\bibitem{algorand}
Yossi Gilad, Rotem Hemo, Silvio Micali, Georgios Vlachos, and Nickolai
  Zeldovich.
\newblock Algorand: Scaling byzantine agreements for cryptocurrencies.
\newblock In {\em Proceedings of the 26th Symposium on Operating Systems
  Principles, Shanghai, China, October 28-31, 2017}, pages 51--68. {ACM}, 2017.

\bibitem{Guerraoui2019}
Rachid Guerraoui, Petr Kuznetsov, Matteo Monti, Matej Pavlovic, and
  Dragos-Adrian Seredinschi.
\newblock The consensus number of a cryptocurrency (extended version), 2019.

\bibitem{guerraoui2022consensus}
Rachid Guerraoui, Petr Kuznetsov, Matteo Monti, Matej Pavlovic, and
  Dragos-Adrian Seredinschi.
\newblock The consensus number of a cryptocurrency.
\newblock {\em Distributed Computing}, 35(1):1--15, 2022.

\bibitem{herlihy1990linearizability}
Maurice Herlihy and Jeannette~M. Wing.
\newblock Linearizability: {A} correctness condition for concurrent objects.
\newblock {\em {ACM} Transactions on Programming Languages and Systems},
  12(3):463--492, 1990.

\bibitem{Hewitt1973}
Carl Hewitt, Peter~Boehler Bishop, and Richard Steiger.
\newblock A universal modular {ACTOR} formalism for artificial intelligence.
\newblock In {\em Proceedings of the 3rd International Joint Conference on
  Artificial Intelligence. Standford, CA, USA, August 20-23, 1973}, pages
  235--245. William Kaufmann, 1973.

\bibitem{lamport1978time}
Leslie Lamport.
\newblock Time, clocks, and the ordering of events in a distributed system.
\newblock {\em Communications of the ACM}, 21(7):558--565, July 1978.

\bibitem{lamport1989paxos}
Leslie Lamport.
\newblock The part-time parliament.
\newblock Technical Report SRC Report 49, Digital Equipment Corporation Systems
  Research Center, Palo Alto, CA, September 1989.

\bibitem{lamport1998paxos}
Leslie Lamport.
\newblock The part-time parliament.
\newblock {\em ACM Transactions on Computer Systems}, 16(2):133--169, May 1998.

\bibitem{Mazurkiewicz1987}
Antoni Mazurkiewicz.
\newblock Trace theory.
\newblock In Wilfried Brauer, Wolfgang Reisig, and Grzegorz Rozenberg, editors,
  {\em Advances in Petri Nets 1986, Part II: Proceedings of an Advanced Course,
  Bad Honnef, 8.--19. September 1986}, volume 255 of {\em Lecture Notes in
  Computer Science}, pages 279--324. Springer, 1987.

\bibitem{nakamoto2008bitcoin}
Satoshi Nakamoto.
\newblock Bitcoin: A peer-to-peer electronic cash system, Nov 2008.
\newblock Accessed: 2025-06-15.

\bibitem{ongaro2014raft}
Diego Ongaro and John~K. Ousterhout.
\newblock In search of an understandable consensus algorithm.
\newblock In {\em Proceedings of the 2014 {USENIX} Annual Technical Conference,
  {USENIX} {ATC} 2014, Philadelphia, PA, USA, June 19-20, 2014}, pages
  305--320, Philadelphia, PA, June 2014. USENIX Association.

\bibitem{Preguica2008CRDT}
Nuno Pregui{\c{c}}a, Marc Shapiro, and Jose~Legatheaux Martins.
\newblock Designing a commutative replicated data type for cooperative editing
  systems.
\newblock Research Report TR-02-2008 DI-FCT-UNL, Universidade Nova de Lisboa,
  Dep. Inform{\'a}tica, FCT, 2008.

\bibitem{ramseyer2024groundhog}
Geoffrey Ramseyer and David Mazi{\`{e}}res.
\newblock Groundhog: Linearly-scalable smart contracting via commutative
  transaction semantics.
\newblock {\em CoRR}, abs/2404.03201, 2024.

\bibitem{Shapiro2007CRDTReport}
Marc Shapiro and Nuno Pregui{\c{c}}a.
\newblock Designing a commutative replicated data type.
\newblock Research Report RR-6320, Institut National de Recherche en
  Informatique et en Automatique (INRIA), October 2007.

\bibitem{shapiro2011crdt}
Marc Shapiro, Nuno~M. Pregui{\c{c}}a, Carlos Baquero, and Marek Zawirski.
\newblock Conflict-free replicated data types.
\newblock In Xavier D{\'{e}}fago, Franck Petit, and Vincent Villain, editors,
  {\em Stabilization, Safety, and Security of Distributed Systems - 13th
  International Symposium, {SSS} 2011, Grenoble, France, October 10-12, 2011.
  Proceedings}, volume 6976 of {\em Lecture Notes in Computer Science}, pages
  386--400. Springer, 2011.

\bibitem{vogels2009eventuallyconsistent}
Werner Vogels.
\newblock Eventually consistent.
\newblock {\em Commun. {ACM}}, 52(1):40--44, 2009.

\bibitem{Yang2024PracticalRateless}
Lei Yang, Yossi Gilad, and Mohammad Alizadeh.
\newblock Practical rateless set reconciliation.
\newblock In {\em Proceedings of the {ACM} {SIGCOMM} 2024 Conference, {ACM}
  {SIGCOMM} 2024, Sydney, NSW, Australia, August 4-8, 2024}, pages 595--612.
  {ACM}, 2024.

\end{thebibliography}

\appendix

% Each appendix section starts on a new page. 
\newpage
\section{Understanding SETL}
\label{sec:understanding-SETL}

A \SETL state includes a finite map from accounts (i.e., addresses) to the account local states.
It also includes a settled claim set that contains all the claims settled so far.
Every settled claim is naturally associated with the unique account that submits and signs it,
and this way the settled claim set can be naturally partitioned by accounts.
Claims will be processed and settled in sequences, called \emph{blocks},
causing the current \SETL state to change.
Depending on the actual content of the block, the state may change in various ways:
a new account may be created,
a new data item may be added to an account,
an existing field of an account may be updated,
some native or non-native (i.e., smart contract) assets may be transferred from one account to another account, and so on.
The settled claim set, on the other hand, will always change in a monotonically increasing way:
it is extended by the said block.

%Claim blocks are issued by accounts, and they are generated by fixed \SETL scripts associated to each account at creation time.
%Through mechanisms imposed with validators, which can be imposed by users or clients of the client in question, and validator-executed \SETL scripts,
%we can assume that accounts cannot generate any other sequence of claims except as prescribed by their scripts.
%Theoretically, this is not a limitation, because as shortly discussed in Section~\ref{sec:creating-accounts}, creating new accounts and issuing claims from those is expected to be easy.
%For example, suppose that $a$ wants to do something new, like voting or buying a new asset, which would require it to execute some new code that it has never executed before.
%Then $a$ can (be required to): (1) create a new account, say $a_{2}$, whose script is the new code (likely a settled claim as discussed in Section~\ref{sec:settleprograms}); (2) add $a$ as the verifier of $a_{2}$ with quorum 1; (3) transfer whatever assets $a_{2}$ needs in order to operate.
%All these will likely be done automatically, with the help of the applications that $a$ interacts with, as well as the issuance of claims from $a_{2}$ and their verification from $a$.
% Consequently, from here on in the paper we only discuss how things can be done and not worry about ``what if the client does something else''.

\subsection{Two Types of Accounts}

We define two types of accounts in \SETL, namely \emph{user-driven} and \emph{contract-based}:
\begin{enumerate}
\item
A \emph{user-driven account} is fully controlled by the account owner, defined as the possessor of its private key, and is not prescribed to any \SETL scripts.
A user-driven account can generate any claims, including but not limited to making native payments, creating accounts,
creating contracts (explained below), creating contract instances, and modifying their account states.
\item
A \emph{contract-based account} is fully controlled by some fixed, prescribed \SETL script that is persistently associated to the account at creation time.
Once created, the script cannot be modified. 
A contract-based account can only submit claims that are generated by the prescribed scripts and thus its behaviors are predetermined.
A contract-based account can only be created by a user-driven account, which does so in order to submit contract-based claims.
\end{enumerate}
\SETL's user-driven accounts and contract-based accounts are similar to Ethereum's externally-owned and, respectively, contract accounts.
Both user-driven and contract-based accounts can receive, hold, and send native tokens and interact with deployed contracts.
Their key differences are also similar to those between Ethereum's externally-owned accounts and contract accounts, listed as follows:
\begin{itemize}
    \item \textbf{User-driven accounts}
    \begin{itemize}
        \item Cost nothing to create;
        \item Can submit any claim block;
        % \item Transactions between user-driven accounts can only be native token transfers;
        \item Are controlled by a pair of public and private keys.
    \end{itemize}
    \item \textbf{Contract-based accounts}
    \begin{itemize}
        \item Incur a creation cost because they are using network storage;
        \item Can only submit claim blocks according to the prescribed scripts;
        \item Respond to messages from other accounts.
    \end{itemize}
\end{itemize}

\subsection{Contracts and Instances}

Both contracts and instances are \SETL accounts and both are contract-based accounts, meaning that their behaviors are predetermined by \SETL scripts.
Their account addresses are their unique IDs, which
are publicly known and thus can be used as references to them.
Contracts and instances both have \emph{fields} and \emph{blocks}.
Fields can hold typed values and they constitute the account states, while blocks are prescribed scripts associated to the contract/instance accounts.
This way, \SETL is similar to an object-oriented (OO) programming language, where contracts and instances are similar to classes and objects, while fields and blocks are similar to class members (fields and methods).
However, as we will shortly see, \SETL's contracts, instances, fields, and blocks differ from their counterparts in OO languages in several important and fundamental ways.

Any users of \SETL, namely the owners of user-driven accounts, can create contracts.
To create a contract, one must provide a set of contract fields and blocks, as well as a set of instance fields and blocks.
Contract fields/blocks are the fields/blocks that belong to the contract.
Instance fields/blocks are the fields/blocks that belong not to the contract, but to its instances.
Contract fields determine the account state of the contract account.
Contract blocks determine the prescribed behaviors of the contract.
Only the contract creator can issue the contract blocks (by signing a special message, discussed shortly).

Once a contract is created, anyone, including the contract creator themselves,
can create instances of that contract.
Each created instance is a separate account and thus has its own separate account state, as
specified by the instance fields of the contract.
The behaviors of each instance are determined by the prescribed instance blocks of the contract.
Only the instance creator, which in most cases is not the same as the contract creator,
may call the instance blocks.

Not all contracts need an instance to work.
A contract can be created for the sake of its contract fields and blocks, in which case it can have
no instance fields and blocks.
Still, one can create an instance of that contract (at a cost), but that instance will have no use,
so it will not be economically wise to do so.
Other contracts expect instances.
For example, a voting contract (see Section~\ref{sec:app-voting}) represents the voting authority,
and its contract blocks specify the voting rules and provide means for controlling the voting,
such as closing it and declaring a winner.
Voters need to create instances of the voting contract in order to vote via the instance blocks,
which only grant voters the right to vote.
Voters cannot, e.g., close the voting early, because they cannot call the contract blocks of the voting contract.
On the other hand, the voting authority cannot vote on behalf of the voters,
nor can any voter vote on behalf of another voter,
because only the instance creator can call the instance blocks from that instance.
Even if the instance blocks are prescribed by the voting authority at the creation of the voting contract, they can only be called by the instance creators.

Via contracts and instances,  \SETL imposes a principled scheme for various parties and stakeholders
to concurrently cooperate and coordinate in a shared time-space.
In this scheme, a contract creator sets the stage by specifying the fields and blocks of the contract,
as well as the fields and blocks of its instances.
From then on, the stage has been set.
The contract account can only behave according to the prescribed contract blocks at the request of the contract creator, while every instance account can only behave according to the prescribed instance block at the request of the instance creator.

A contract creation by a user-driven account 
effectively creates an actor.
The semantics of \SETL only allows the contract owner to send messages to the contract actor.
Indeed, in \SETL, \verb|instance| fields form the state of the contract instance actors, and these instance actors can only initiate \verb|instance| blocks according to the contract.
The fields which do not have an \verb|instance| modifier form the state of the contract actor, and the blocks which do not have an \verb|instance| modifier can only be generated by the contract owner.
In \SETL, anybody can create instances of the contract,
and each instance is an actor controlled by the its owner user-driven account. 
This way, the semantics of \SETL is faithful to the actor model in that actors only communicate with actors they have knowledge about.

A clarification is needed with regards to the fact that owners of contracts and of instances, respectively, can sign on their behalf.
In Section \ref{sec:fastset}, FastSet stipulates that each account $a$ can only submit claim blocks which are signed by $a$.
In \SETL, on the other hand, each contract creation and each instance of it becomes an actor, which has its own state and address.
However, it would be rather inconvenient and likely insecure and unmanageable by users to allow each actor to own its own private key to sign all its claim blocks.
Instead, \SETL allows the contract-based actors to only be controled by their owners, technically just other accounts that have a private key and thus can sign messages.
That is, the unique owner of the created contract, or the unique owner of some instance of it, is the only entity that can sign blocks on behalf of the address corresponding to the associated contract or instance actor.
This is not a modification of the FastSet protocol, but rather a particular implementation of it.
Indeed, a single private key can be used to generate infinitely many public keys through various methods, like hierarchical deterministic (HD) wallets or different cryptographic algorithms.
These different public keys are deterministically derived from the same private key and are used as contract-based actor addresses in \SETL, although there is only one initial pairing between a private key and its primary public key.

\subsection{Templates and Parameters}
\label{sec:templates-and-parameters}

Both contracts and instances are \SETL accounts whose behaviors are determined by prescribed blocks
as \SETL scripts.
In the simplest setting, a user can always liberally create contract-based accounts
with the intended scripts.
However, it turns out that many contract-based accounts are expected to reuse the same scripts,
and only differ in certain parameters, arguments, placeholders, and so on, which may be
instantiated at account's creation time.
To avoid code duplication and save validator state space, \SETL allows \emph{templates}.

Templates are data that can be encoded in the account messages and stored in the account states.
They will be processed by the FastSet validators, possibly with concrete parameters and arguments,
to produce actual scripts and claim blocks, at account creation time or at block generation time.

Templates can be for anything, from simple facts and statements, to books, pictures, videos, programming language specifications, and programs in any programming languages.
FastSet gives them no intrinsic meaning except for their parameters, which can be instantiated.
Applications and/or potential FastSet commands are free to give them any meaning, or no meaning at all.
For example, \SETL will shortly introduce a claim construct called \texttt{contract}, which will create a contract from a template.

Templates are related to macros and, more broadly, to meta-programming.
%They tell validators how to generate the actual, concrete scripts and blocks when needed.
%That is, templates are code generates code.
%
Macros and meta-programming languages are, however, notoriously difficult to design.
Numerous questions have to be answered.
What parameters are allowed to appear in a template?
Should templates have types?
When is the well-formedness and well-typedness of a template checked, statically or dynamically?
Are nested templates allowed?
How and when is a template encoded and decoded?
Since \SETL is meant to be a scripting language, its template support is minimal---introduced by means of examples in Sections~\ref{sec:templates} and \ref{sec:parametric-templates}.
More advanced and meaningful code compositionality mechanisms can be deployed separately on top of this minimal infrastructure, similarly to how programming languages (Section~\ref{sec:pls-vms}) and programs (Section~\ref{sec:settleprograms}) are settled.

\begin{comment}
It is not our focus to answer all the above questions in this section.
Instead, we prefer to take the same approach as how we have designed $\mathbb{K}$ and its frontend language to design templates.
That is,
we start with numerous concrete and actual examples to illustrate how templates should be used,
in a reasonable way,
without bothering too much with formal specifications or implementation details.
After we understand all the examples and reach to an agreement on how templates should be used in those examples, we start to nail down its specification along with the other components of \SETL.
\end{comment}

\subsection{Claim Block Assumptions and Non-Assumptions}

Here we discuss the assumptions and non-assumptions
pertaining to claim blocks in the examples found in the subsequent sub-sections.

We assume that all blocks are deterministic.
We also assume that a block is processed as an atomic action.
Different blocks accessing the same variables will not interleave with each other.
However, sophisticated FastSet implementations that aim at high performance can use synchronization techniques such as locks
to enable parallel execution of independent and non-conflicting blocks.

Unfortunately, we need to compromise the idea of enforcing the claim weak independence assumption made in Assumption~\ref{assumption:addresses}
at the syntactic and language level.
That is to say, using the \SETL scripting language, one is able to
create accounts whose claim blocks are not weakly independent.
This situation happens mostly among a contract and its instances.

For example, in the Dynamic Multi-Sig Account example in Section~\ref{sec:dynamic_multisig},
an instance claim that is currently valid may become invalid
if the contract lowers its \verb|quorum| or removes some verifiers.
Another example is the voting example in Section~\ref{sec:app-voting}.
As explained there, if the voting authority tries to count the number of registered/voted voters
and then take according actions that affect the voters' claims/blocks---such as closing the voting
as soon as it collects 1000 votes---it breaks the weakly independence claim assumption.
Indeed, suppose the authority currently collects 999 votes. It is still possible and valid
to process Alice's vote, or Bob's vote, but not both of them.

We have decided to give clients and applications the freedom
to adventure beyond the safe territory of weak independence.
Some applications, such as dynamic multi-sig or voting, can choose to do so, knowing that
by breaking the weak independence assumption their applications and accounts may get stuck,
in which case FastSet implementations may provide means for them to unstuck themselves by clearing the different versions of \textit{pending} claims among all the FastSet validators.
Other applications which want to stay in the safe domain of weak independence should be incentivised
to use static analysis or formal verification techniques to prove that it is indeed the case.
FastSet/\SETL will also be able to provide a set of guidelines that help to write applications that respect the weak independence assumption.

Given that said, we do expect that situations where the weak independence assumption must be compromised, to be quite rare.
The majority examples presented in the subsequent sub-sections, especially
verifiable computing in Section~\ref{sec:verified-computations},
escrow in Section~\ref{sec:escrow},
and non-native assets and payments in Section~\ref{sec:nonnative_assets},
all satisfy the assumption.

% Each appendix section starts on a new page. 
\newpage
\section{Modifying Settled Data}
\label{sec:database}

In Section~\ref{sec:settle-data} we showed how to settle persistent data and records
on FastSet, by creating an account whose fields are initialized to the data/records
that should be settled. 
Since the fields cannot be modified after initialization, the weak independence property
holds. 

Here we discuss possible extensions where the settled data can be modified without
compromising the weak independence property. 
In particular, as long as the FastSet validators do not allow any other account 
(except the account that owns the settled data) to
generate claims that directly read the values of the settled data,
the weak independence property holds. 
The other account can still read the values of the settled data---the FastSet validators
may actually provide SDKs for it---and generate different claims depending on the values they read.
However, they cannot generate claims whose behaviors depend on those read values. 
Any read-value-sentitive decisions must be made by the clients outside FastSet,
and not by the FastSet validators. 

To allow a data owner modify their settled data, we only need to define some setter blocks. 
For example, if Alice wants to update her name, she can add a new setter block:
\begin{verbatim}
    set_name(new_name) {
        verify({gov}, 1);
        name := new_name;
    }
\end{verbatim}

The data owner can also generate conditional claims about their settled values. 
Let us add a new field \verb|age| to the account \verb|ALICE_INFO|.   
Then, Alice can claim that her name is Alice Wonderland and her age is above 21, by generating the next claim:
$$
  \langle \verb|ALICE_INFO.name == "Alice Wonderland" and ALICE_INFO.age > 21|
  \rangle_\texttt{alice}
$$
We tacitly assume such boolean constructs, using standard notations (e.g., \texttt{==}, \texttt{>=}, \texttt{and}, etc), noting that FastSet implementations will likely charge clients fees that are a function of the complexity of evaluating such boolean expressions---since the validators will eventually have to evaluate them.

The weak independence property holds for \verb|ALICE_INFO| because all the 
reads (e.g., \verb|ALICE_INFO.name|) and writes (e.g., \verb|set_name()|)
take place under one account \verb|alice|, so they are totally ordered. 
The other accounts can only look at the claims that Alice settled
and they must be careful with how to interpret the claims. 
Indeed, the claim 
$$
  \langle \verb|ALICE_INFO.name == "Alice Wonderland" and ALICE_INFO.age > 21|
  \rangle_\texttt{alice}
$$
only shows that \emph{once}, 
Alice has/had such name and such age, but it does not say that
\emph{at present}, Alice has such name and such age.
Any attempt to achieve the latter will break the weak independence property
because it is equivalent to allowing reads and writes of shared locations, which are not weakly independent.
Even if the FastSet implementation provides the capability to clients to query other accounts' local fields, say through an API, they should not rely on the values of those queries.
For example, it may be that Alice has submitted her name change claim, but all validators happen to have that claim presettled (in Step 6 in Figure~\ref{fig:fastset}) but not settled yet (in Step 7 in Figure~\ref{fig:fastset}), so the name query made to any validator would still return the old, incorrect name, even if made by Alice herself.
Alice can settle the query herself \textit{as a claim} only after the validators have settled all her previous claims, so her query claim can be trusted to be correct.

%Even if validator API access to accounts data might be allowed as a service, FastSet implementations should \textit{not allow clients to make claims about other accounts local variables}.
%Doing so may lead to non-determinism and accounts frozen in pending status (in Step 4 in Figure~\ref{fig:fastset}).
Other clients can also ask Alice to re-claim her name and/or age, to prevent Alice from abusing
a settled but deprecated and invalid claim.
For example, Alice may settle a claim saying that her name is ``Alice Wonderland'' and then completes a name change,
but the previous claim has been settled for good.
In this situation, other clients can use timestamps (explained in Section~\ref{sec:timestamping}) and reject any claims that they consider as too dated to be relevant, at the cost of potentially breaking the weak independence property.

\end{document}